# It Takes Three to Ceilidh: Pension System and Multidimensional Poverty Mitigation in China

Yansong David Wang[a], Tao Louie Xu[b], and Cheng Yuan[a]

*[a] Business School, Nanjing University, 22 Hankou Road, Nanjing, 210046, China*

*[b] School of Social and Political Science, The University of Edinburgh, 15a George Square, Edinburgh, EH8 9LD, UK*

**Author's Note**

Yansong Wang, Email: yansong.wang@smail.nju.edu.cn, ORCID: https://orcid.org/0009-0007-9700-7063

Tao Xu, Email: s2475073@ed.ac.uk, ORCID: https://orcid.org/0000-0003-0510-3343

Cheng Yuan, Email: yuanch@nju.edu.cn, ORCID: https://orcid.org/0000-0003-2027-9064

# It Takes Three to Ceilidh: Pension System and Multidimensional Poverty Mitigation in China

This research, employing the Alkire-Foster approach to uncover multidimensional poverty between 2012 and 2020 in China, models and examines the sustainable effects and mechanisms of the three-pillar pension system in household poverty mitigation with the China Family Panel Studies data. The results evince that more participation in the pension system mitigates the probability of being trapped in multidimensional poverty. The findings reveal the significance of synchronising state social insurance, enterprise annuity, and individual commercial insurance. The mitigation effect of market-oriented pillars is achieved through more investment in and consumption for livelihood assets. Based upon the sustainable livelihoods framework, livelihood assets ameliorate household capabilities in human, natural, financial, and psychological capital against exogenous shocks and uncertainties. Our research contributes to the theory and praxis of the facilitating state in new structural economics illustrated by the multi-pillar pension system in sustainable micro-household poverty mitigation.

***Keywords:*** multidimensional poverty; three-pillar pension system; facilitating state; sustainable livelihoods; new structural economics; China

## 1. Introduction

With an acceptable income, do you still worry about access to high-quality education? Do concerns arise over your family's health? Or perhaps you watch friends or relatives striving for decent work and a stable life? If not, you are fortunate; if the questions resonate, then you too are entangled in the experience of multidimensional poverty (MP).

MP remains a global concern, pervasive even as income-focused policy intervention in the Global South has mitigated absolute poverty by elevating household incomes. Yet, the income-based unidimensional metrics alone obscure the entrenched social disparities and inequalities in education, health, and decent work (Sen, 1999; Alkire & Santos, 2014). In praxis, income-focused intervention conquers destitution but fails to integrate marginalised groups into sustainable livelihoods with multidimensional indicators, such as healthcare and education. Apropos the metrics, emerging multidimensional indicators highlight inequalities otherwise hidden by unidimensional measurement (Alkire & Santos, 2014; UNDP, 2010). For instance, the United Nations Development Programme's (UNDP, 2023) Multidimensional Poverty Index (MPI) reveals that over a population of one billion—more than a fifth of that all around the world—are grappling with MP. Measuring these concerns, uncontrollable global health crises have reversed progress in MP mitigation by about one decade (UNDP, 2020). The worsening trend of population ageing worldwide, meanwhile, further aggravates health disparities and general conditions for households with limited income and education. Population ageing deepens social inequalities and unsustainable livelihoods, particularly in the South (Chen et al., 2022). The combined effects of the concerns undermine efforts against inequalities but entrench vulnerabilities within marginalised communities.

The multidimensional intervention targeting older persons is pragmatic for household MP mitigation. The higher costs of oldercare increase the MP probability among older adults and their households (Duflo, 2000; Barrientos, 2003; Case, 2004). The costs strain household resources, diverting money from education, nutrition, and care; this financial scarcity affects younger generations who sacrifice savings to support oldercare. When oldercare is being unaffordable, older persons are required to work into their seventies and even eighties due to meager pensions, while younger generations bear the dual burden of the responsibility to support older adults and concerns about their own care. Accordingly, the oldercare burden restrains intergenerational expenditures on sustainable livelihoods, triggering an investment decline in education, healthcare, and so forth at the household level. Besides, households suffer from uncertainty and an increased likelihood of unexpected exogenous shocks in the post-pandemic era of economic volatility and geopolitical tensions. Resolving this burden with a well-designed pension system may mitigate household poverty sustainably in society. Since the pension system serves as a key to the oldercare burden at individual, household, and national levels, this research centres on the social policy of pension provision and its anti-MP implications from the perspective of state-market tango.

To settle the unsustainability and inequality, it takes two to tango—between the state and markets—around which policy debates are framed and policymaking principles diverge (Gottschalk & Danziger, 1985). The tango simile inspires poverty mitigation solutions with selective state intervention, not limited to pro-growth policy (Gottschalk & Danziger, 1985) and aid transfer (Haushofer & Shapiro, 2016). Amidst the inspiration from Lin's (2012) new structural economics, the social policy tango emerges as where the role of facilitating state in inclusive development operates. The social policy targets poverty and inequality mitigation through a nuanced grasp of demographic structures and individual characteristics. Initially, Lin guides industrial policy by aligning with factor endowments and economic structures, but the principles can extend to social policy by recognising the ageing population, expenditure patterns, and income inequalities as structural determinants of social policy. Hereto, a facilitating state must not only respond to the demographic realities but also create enabling conditions for households to reduce precautionary savings, invest in healthcare and education, and achieve sustainable livelihoods. Integrating new structural economics into social policy, since MP mitigation is achieved through strategic coordination of state facilitation and market mechanisms, can tailor policymaking to the structural contexts of demographics. China's contribution to the human struggle against poverty is a result of combining its facilitating state with effective markets (Lin, 2012). With selective neoliberal reforms since 1978, China has liberated over 800 million people from extreme poverty. China's facilitation brings a reduction in the global poverty rate from 44% to 9% (World Bank, 2018); from 2010 to 2014, the state lifted 69 million people out of MP (UNDP, 2023). With the discourse of facilitating state, the evolution of China's social policy generates a compelling case to grasp the tango between state-directed and market-driven measures in poverty mitigation.

It takes two to tango, but three to ceilidh. With the state-market tango, our other simile portrays the three-pillar pension system (TPPS) as the ceilidh. Ceilidh is a form of group dancing with three or more participants, necessitating coordination among actors. Similarly, the TPPS spreads the risk and shares responsibility across diverse pension pillars and market actors. Our research argues for a three-pillar systematic arrangement wherein the TPPS mirrors the ceilidh dancing. Although Lin's conceptualisation of facilitating state inspires policy intervention to be selective and proactive, it is regrettable that few studies value the pension system as a facilitative intervention of social policy. The channels and mechanisms could only be inferred from pension studies literature. Thus, focused on how market-oriented engagement can reconcile with the imperative of social security, this research explores how China's TPPS navigates policy intervention and the invisible hand. Analysing the causality of the TPPS in mitigating poverty contributes to academic debates and real-world policymaking. Besides, it uncovers the effects and mechanisms by examining the relationship between TPPS and MP, so as to inform social security policymaking in poverty mitigation and consolidate our understanding of the interplay among the facilitating state, markets, and individuals in inclusive development.

China's TPPS derives from a framework designed to settle the pension challenges by sharing responsibility among the state, enterprises, and individuals (World Bank, 1994). The first pillar is a government-managed, defined-benefit mandatory state social insurance as a component of China's basic social security. The second one consists of a privately-managed, defined-contribution annuity, established by enterprises or relevant organisations to reduce people's reliance on the first pillar through accumulated savings. The third now is voluntary commercial insurance invested by individuals. The TPPS ceilidh mitigates the pressure on state pension funds and diversifies the sources (Barr, 2002; Sørensen et al., 2016; Natali & Raitano, 2022). For instance, amid the stage of stagflation and recession since the 1970s, the multi-pillar systems successfully dealt with the sustainability crises of public finance around pensions in Chile, Argentina, and others (Holzmann et al., 2000). The TPPS not only yields premium returns (Gollier, 2008) but facilitates risk-sharing across generations (Broeders et al., 2021; Chen et al., 2023). In praxis, multi-pillar systems are advocated by the OECD (1998) and the European Parliament (2015). Their advocacy argues for a balanced ceilidh-like system of the state and markets. China adopts it: state social insurance, enterprise annuity, plus individual commercial one. However, China's TPPS appears imbalanced as the first pillar, to date, remains dominant, whereas the other two are characterised by uneconomical payment and limited coverage. Therefore, promoting TPPS participation is crucial to mitigate household MP, yet a gap exists in the evaluation of the effects and mechanisms of the TPPS and those of each pillar in MP mitigation.

The limitations of existing knowledge emerge as available literature downplays the TPPS participation and the identification of participants. Most empirical studies merely use the collection as the key variable, uncovering the individual effects of collecting pensions (Maitra & Ray, 2003; Case, 2004). For example, participants in health insurance immediately benefit from returning healthcare, whereas pension insurance involves a decades-long gap and financial costs between participation and collection. Solely collecting the evidence of pension collection contributes to neither financial nor development economics. By contrast, our TPPS participation cares about policymaking and public welfare. Furthermore, even when discussing pension participation, most examinations treat it as a binary variable. This dichotomy ignores the specific pillars and fails to elucidate their particular effects. Additionally, previous studies rarely differentiate participants by age, highly confusing those participating in and collecting pension insurance within the sample.

The costs of TPPS participation concern younger participants since the cumulative expenses increase with multi-pillar engagement. Although higher expenditures might raise a concern about the anti-MP efficacy of TPPS, we present a simple overlapping-generations model that offer a response. It also shifts consumption and investment patterns by increasing future income expectations (Carroll, 1994; Adams & Rau, 2011). The sustainable livelihoods framework further embeds our MP mitigation in human development and resilience, aligning with Sen's (1999) capability approach to settle multidimensional deprivation (Scoones, 1998; DFID, 1999; Natarajan et al., 2022). Hence, this research aims to offer a more robust causal inference between TPPS and MP, informing the causality and evidence for policymaking.

The remainder is structured as follows. Section 2 reviews the background of the TPPS

provision designed by the facilitating state of China and the MP situation. Section 3 offers a simple model to demonstrate participation in TPPS mitigates MP, followed by the mechanisms with reference to the sustainable livelihoods framework. Section 4 designs the empirical strategy. With our results, Section 5 indicates that a state-market coordinative three-pillar pension provision is facilitative to China's poverty mitigation, and then Section 6 examines the mechanisms for sustainable MP mitigation by multiple pillars relevant to the livelihood assets. Section 7 concludes with our discussion and implications.

## 2. Background

### 2.1 Three-Pillar Pension System by Facilitating State

China's pension system has reformed from a state-dominated responsibility framework to a state-market coordinated responsibility-sharing system, facilitated by the facilitating state, despite long-standing criticism and skepticism. Initially, China's pension system commenced in 1951 with the Basic Pension Insurance System, which was financed by the state and implemented by enterprises (Dixon, 1981). For Communist China, enterprises were either state-owned or collectively-owned, and the Basic Pension Insurance System primarily covered employees in state-owned and some collectively-owned entities. In other words, the single-pillar system basically achieved near-universal coverage of the working class within the communist framework. However, due to the colonial legacy and the devastation of wars, the People's Republic inherited extreme poverty and meager state revenues. As a result, a non-contributory, state-financed pension system was unsustainable. Beyond sustainability concerns, the state-directed system also aggravated inequalities. For instance, the single-pillar system neglected welfare protection for the peasant class. Consequently, to reduce inequalities in social welfare provision, China adopted the role of facilitating state, to reform the system by sharing the responsibility across multiple pillars and involving market actors in the process (Liu & Sun, 2016).

[Insert Figure 1 here]

In 1978, the government of China adopted selective neoliberal reforms, marking what is seen as the beginning of the praxis of new structural economics (Lin, 1992). Following the concept of the facilitating state in new structural economics, China judiciously expanded the coverage of its pension system and facilitated market mechanisms. Lin's new structural economics, initially deployed to analyse industrial development, spotlights the endogeneity and dynamism of the economic structure in transformation (Lin, 2012). It advises the state to consider factor endowments and structural conditions at each development stage, judiciously facilitating the role of markets rather than a one-size-fits-all shock therapy seen in other states. Similarly, in the realm of social policy, inherent endowments and structures are also crucial (Xu, 2024). Just as industrial policy aims to build competitive advantages, the goal of social policy is to mitigate poverty and inequality. Specifically, a facilitating state must account for inherent demographics such as population structure and individual characteristics of people's consumption and investment, facilitating the multi-pillar system's efficacy in inequality and poverty mitigation. Under neoliberal reforms, state-owned enterprises were given autonomy over their finances, but in urban areas, the large number of retired workers placed a heavy burden due to the unfavourable age structure, making it impossible for enterprises to bear the high pension costs. Therefore, to cope with the consequences of economic transition, the social pension system was first implemented in urban areas (Feldstein, 1999; Xu, 2022). Regrettably, the neoliberal pension system reforms diverged between rural and urban areas, exacerbating the inequalities in social welfare provision in transition until the 2000s.

Fortunately, drawing on the lessons of selective neoliberal reforms, the facilitating state of China established a pension system with three pillars in 2004. The TPPS is dominated by public pension schemes, supplemented by enterprise annuity and individual commercial insurance. It reflects a policy praxis that combines the role of the facilitating state with effective markets. Since the 1990s, the state shifted the responsibility of pension provision to be shared by the state, enterprises, and individuals, aligning with the global pension reform trends advocated by the World Bank (1994). Thanks to the facilitating state's role, China's TPPS has been adjusted based on the specific endowments and demographic structures, as well as individual characteristics relevant to economic behaviours. For instance, China's partial funding scheme aligns with demographic shifts caused by the one-child policy (Liu & Sun, 2016). This scheme implies that the future expected pension income of households participating in the first pillar is composed of two parts: 1) future pension contributions based on the pay-as-you-go principle (social pooling account), and 2) the accumulation from individual participation (individual account). Conversely, a fully pay-as-you-go system is highly susceptible to demographic changes. The pay-as-you-go means that one generation's contributions are directly used to pay the pensions of the currently retired generation. If households participating in the pension system expect a future decline in population and income, this insecurity may reduce their investment in and consumption for livelihood assets. On the other hand, a fully individual full-funding contribution-based scheme would exclude poorer households and exacerbate inequalities.

Additionally, the market risks and uncertainties brought about by economic transition have driven households to blindly pursue security through precautionary savings and high savings rates (Chamon & Prasad, 2010), crowding out household expenditures on education and healthcare. Unlike a fully marketised and financialised pension system, the facilitating state of China advocates the enterprise annuity policy, offering tax incentives to enhance the sustainability of the TPPS, particularly the security given by the second and third pillars. This sense of security reduces the need for precautionary savings and increases expenditures on livelihood assets.

Specifically, rural areas experienced mixed social and economic impacts from the reforms. With the implementation of the one-child policy and population migration driven by urban industrialisation, traditional family-based oldercare became increasingly insecure and unsustainable. In 1999, China began to establish the *Rural Pension System*, but compared to urban areas, the individual full-funding contribution scheme overlooked the urban-rural disparities, the income inequalities driven by demographic structures, and the lack of trust among the peasant class in the pension system (Shi, 2006). Hence, referencing the principles of new structural economics, in 2009, China's facilitating state intervened to moderate the neoliberal reforms through substantial investment in social security (Liu & Sun, 2016). The investment established the *New Rural Social Pension System*, enabling local governments to use tax revenues to finance rural residents, handling the shortcomings of the individual full-funding scheme. Additionally, government subsidies were introduced to encourage individual participation. After 2014, the Urban Residents' Pension System and the *New Rural*

*Social Pension System* were gradually merged into a unified *Urban and Rural Residents' Social Pension Insurance System*. The new integrated system reduced urban-rural inequalities in social welfare provision, providing basic security for the majority of people (Cheng et al., 2016; Liu & Sun, 2016; Xu, 2022).

### 2.2 Multidimensional Poverty in China

Poverty in China was entrenched. Since the founding of the People's Republic of China in 1949, the state engaged in relief-based anti-poverty policy, transferring material and financial aid to impoverished groups in rural and hinterland areas for their basic livelihoods. After the neoliberal reforms initiated in 1978, China has achieved its respective milestones in poverty mitigation (World Bank, 2018; Alkire & Fang, 2019). China's substantial progress in reform and development has eradicated absolute poverty across urban and rural regions. The World Bank (2018) reports that China, from 1978 to 2015, extricated 850 million residents out of extreme poverty, measured by an income of $1.90 per day. Although it is commendable, there is no other means to sustainable development than considering the multidimensional indicators underlying impoverishment.

The demographics of excessive precautionary savings and an ageing population exacerbate multidimensional indicators of impoverishment among Chinese households. Notably, high saving rates suppress macroeconomic growth (Mody et al., 2012), yet the literature has not adequately examined their micro-level effects. For instance, the perceived security gained from high savings prevents residents from investing and consuming against their MP vulnerability (Chamon & Prasad, 2010). Assuming fixed income, it is rational to infer that savings and expenditures on consumption and investment are substitutes. High saving rates constrain household expenditures on education, healthcare, and living standards, curbing human and psychological capital development (Luthans et al., 2004; Goldin, 2016; Xu et al., 2021). Concerns about uncertainty and risk often result in reduced expenditures, which can mean forgoing further education, nutritional healthcare, recreation, and salubrious habitation for the next generations. Despite the belief that high savings act as a buffer against shocks, inflation risk is frequently overlooked; high saving rates can incur overly conservative financial decision-making and limited income diversification due to meager financial capital. The declining real purchasing power of money also erodes confidence in consumption and investment.

In some cases, households liquidate financial and natural assets in favour of precautionary savings, weakening livelihood assets with adverse impacts on MP. Additionally, an ageing population intensifies household concerns about future uncertainties, increasing the burden on family expenses and fuelling their precautionary savings (Chamon & Prasad, 2010; Chen et al., 2022). These demographic factors and structural characteristics underscore the relevance of new structural economics in social policy: social policy must consider the endogeneity and heterogeneity of population structure and individual characteristics of consumption and investment to craft pro-sustainable livelihoods poverty-reduction policy.

The facilitating state of China, according to new structural economics, has recognised the limitations of any unidimensional approach and increasingly values compulsory education, primary healthcare, secure housing, and clean water to which hinterland and rural regions were deprived of equal access. Following the triumph of poverty mitigation, emerging challenges manifest in China's struggle against MP in rural and hinterland areas. The evolution of China's approach to poverty mitigation unfolds a shift from a relief-based to a development-based strategy. The state was distributing immediate relief through food and cash aid. Step by step, China's social security plays an increasingly crucial role in reducing MP. For instance, China's social security system includes a range of schemes aimed at oldercare, healthcare, and unemployment. Among the security systems, the pension one has undergone significant reforms, adaptive to the ageing population and transitioning from state-controlled single-pillar security to TPPS-coordinated sustainability. The reforms are instrumental in ensuring the economic security of the older persons because the state was long responsible for their decent life after retirement. From 2010 to 2014, 69 million people in China, especially older individuals, escaped from MP, making a significant contribution to the global sustainable poverty mitigation as a reference for other South states.

Using China Family Panel Studies (CFPS) data, our research measures and decomposes the multidimensional poverty in China from 2012 to 2020. Above all, it is necessary to consider the dimensions and their indicators. Our measurement is inspired by the Alkire-Foster approach and MPI—which not only reflects the dimensions of the Human Development Index but also includes ten dimensions from the United Nations Millennium Development Goals. MPI has been recognised by states with different political institutions and ideologies, and its robustness works well (Pham et al., 2020), expanding the practical value of the Alkire-Foster approach (Alkire & Santos, 2014). Therefore, the three dimensions of health, education, and living standards are selected. Considering the indicators, available literature based on China's conditions and contexts, including Alkire and Fang (2019) and Liu et al. (2023), contributes to our selection which is as Table 1.

[Insert Table 1 here]

Based on the indicators and weights established above, our research employs the Alkire-Foster approach (see Appendant Notes) to calculate the the year-on-year decrease in MP incidence among Chinese urban and rural households across different multidimensional cutoffs of poverty denoted by $k$ (Figs. 2 and 3). When $k$ exceeds 0.6, the incidence approaches zero for both urban and rural areas. This indicates that as the poverty cutoff becomes higher, the relative anti-poverty improvement features more significant, following China's endeavours to address extreme poverty and the efficacy of targeted poverty reduction policy. However, the urban-rural disparities remain evident, with urban regions consistently showing lower poverty incidence across all $k$-values compared to rural areas. Additionally, the concavity and convexity of the figures indicate that the rate of MP reduction in urban areas is more effective than in rural areas, which is in stark contrast to the conclusions drawn from the commonly used income-based poverty measurements. Such results must be taken seriously, as a lower poverty index and faster MP reduction might further widen the urban-rural gap. This disparity underscores the need for continued dedication in rural development and fair resource distribution towards sustainable MP mitigation.

[Insert Figure 2 and Figure 3 here]

Drawing on the UNDP standard, the MP cutoff $k$ is set at 0.33 (Alkire & Foster, 2011; see Table 2), allowing for further identification of multidimensionally poor households. The data from 2012 to 2020 reveals a consistent decline in poverty incidence across all categories, reflecting China's significant strides in poverty reduction. Notably, the decline shown in figures is more significant for urban households, suggesting that urbanisation-style development is effective in reducing poverty. The rural poverty rate, while higher initially, also demonstrates a substantial decrease, indicating China's efforts of anti-poverty policy.

[Insert Table 2 here]

The H index, which measures the proportion of multidimensionally poor populations, reflects the overall reduction in poverty. The A index is average deprivation share, capturing the degree of poverty experienced by households. This is particularly evident in the rural regions. The metric $M_0$ is of interest as it encapsulates both the MP incidence and degree. The decline of $M_0$ across the entire sample aligns with expectations. The reduction in the $M_0$ implies the development policy efficacy in not only reducing the number of poorest population but improving people's living conditions. However, an anomaly arises in rural areas, which calls for further investigation. Upon incorporating the income dimension into our recalculations, it is observed a consistent downward trend in $M_0$ for all areas (Appendant Figure 2). This observation reinforces the necessity of multidimensional measurement. While economic growth in a country is likely to contribute to improvement in residents' education, health, and living standards, this is not always guaranteed especially considering disparities between different regions. For a more detailed review of the causes of MP, it is imperative to dissect the various dimensions (Tables 2 and 3).

[Insert Table 3 here]

Table 3 presents the contribution of each dimension to the poverty index at the specified value of $k$. Taking 2012 as an example, $k$ =0.33 is MP cutoff, and $k$ =0.66 is extreme MP cutoff. When $k$ =0.33, health contributes the most significantly at 40.1%, followed by education at 38.5%, with living standards accounting for the remaining 21.4%. By adjusting the MP cutoff $k$, the contribution of each dimension to MP has not seen particularly significant changes. Health remains most prominent and is deteriorating year by year. More concerning is that, compared to rural households, urban ones contribute a larger proportion to health poverty. It may be attributed to the adverse consequences of economic growth, such as environmental pollution, intense labor, and increased risk of infectious diseases. This highlights the limitations of using income to measure poverty.

### 3. Theoretical Framework

Pension collection demonstrably supports poverty mitigation (Duflo, 2000; Barrientos, 2003; Maitra & Ray, 2003; Case, 2004). Regrettably, there remains a dearth of literature examining the specific impact of pension insurance participation on MP. Intuitively, participation in pension system might impose an additional financial burden, potentially crowding out household expenditures on education, healthcare, and living standards, thus aggravating MP. However, when shifts in savings, expectations, and intertemporal household decisions are considered—this presumption warrants scrutiny (Carroll, 1994; Hubbard et al., 1995; Bruhn and Steffensen, 2011; El Mekkaoui De Freitas and Oliveira Martins, 2014).

### 3.1 Model

**Without TPPS.** Assume that individuals live for two distinct periods—younger and older. In period $t$, a younger individual $i$ denoted by subscript 1 allocates the income $W_{it}$ between expenditure $E_{i1t}$ and savings $S_{it}$. In period $t$, an older individual $j$ expenses the entirety of their remaining wealth $E_{i2t}$. $E$ implies not only everyday consumption $E^C$ but also the acquisition of various livelihood assets $E^I$. For simplicity, $E^C$ is provisionally assumed as a constant. The individual utility function is:

$$U_{it}(E^I_{i1t}, E^I_{i2t+1}) = LnE^I_{i1t} + \frac{1}{1+\rho} LnE^I_{i2t+1}, \rho > -1, i = 1,2,3..., j = 1,2,3... \quad (1)$$

Suppose the government levies a proportional tax at rate $q$ ($0 < q < 1$), then the expenditure budget constraint is:

$$E^I_{i1t} + S_{it} = W_{it}(1-q) \quad (2)$$

$$E^I_{i2t+1} = S_{it} + S_{it}r_{t+1} \quad (3)$$

$$E^{I*}_{i1t} = \frac{W_{it}(1-q)(1+\rho)}{2+\rho} \quad (4)$$

where $r_{t+1}$ is the real interest rate in period $t+1$. It solves the utility maximisation problem subject to the constraint.

**With TPPS.** Assume that in period $t$, a younger individual $i$ participates in the first pillar of the pension system. As detailed in Section 2, the younger individual's social pension account consists of the social pooling account and the individual account. The social pooling account is funded by employer contributions, which are mandatory benefits that require no personal contributions and pay a return after retirement. The individual account includes individual contributions $p_a$ and government subsidies $s_t$. Contributions to the second and third pillars are $p_b$ and $p_c$, respectively. $p_a$, $p_b$, and $p_c \geq 0$ and not all of $p_a$, $p_b$, and $p_c$ are zero as one participates in at least one pillar. Assuming the return rates of pensions from the first, second, and third pillars are $g_a$, $g_b$, and $g_c$; the utility function remains, the expenditure budget constraint is as follows.

$$E_{i1t}^{I} + S_{it} = W_{it}(1-q) - p_a - p_b - p_c \quad (5)$$

$$E_{i2t+1}^{I} = S_{it} + S_{it}r_{t+1} + (1+g_a+s)p_a + (1+g_b)p_b + (1+g_c)p_c + MB \quad (6)$$

$$E_{i1t}^{I**} = \frac{W_{it}(1-q)(1+\rho)}{2+\rho} + \frac{(1+\rho)[(g_a+s-r)p_a + (g_b-r)p_b + (g_c-r)p_c + MB]}{(2+\rho)(1+r)} \quad (7)$$

**TPPS Participation and MP Mitigation.** The change in livelihood assets expenditure with participation in TPPS as compared to their absence is:

$$\frac{E_{i1t}^{I**} - E_{i1t}^{I*}}{E_{i1t}^{I*}} = \frac{[(g_a+s-r)p_a + (g_b-r)p_b + (g_c-r)p_c + MB]}{W_{it}(1-q)(1+r)} \quad (8)$$

In addition, the Multidimensional Poverty Index $MP_i$ is a strictly decreasing function $F$ of livelihood assets expenditure $E_{i1t}^{I*}$.

$$MP_i = F(E_{i1t}^{I*}) \quad (9)$$

The increase or decrease in expenditure is correlated with the rates of return on each pension pillar and the real interest rates on savings. By 2023, China's first pillar pension fund reported an average annual return of 5.00%, while the second pillar achieved an average annual return of 6.26%. The third pillar's returns are determined by the operations of commercial insurance companies. Each pillar's rate of return significantly exceeds the risk-free rate. Additionally, factors $s$ and $MB$ merit attention. There is an inverse relationship between livelihood assets expenditure $E_{ilt}^{I*}$ and $MP_i$. Consequently, it can be deduced that participation in either the first, second, or third pillar has a positive effect on MP mitigation. The more pillars—the more quantity of non-zero values for $g_a$, $g_b$, and $g_c$—the lower the $MP_i$. Besides, the increase in livelihood assets expenditure is inversely proportional to $W_{it}$. That is, under ceteris paribus conditions, the lower the real wages of a younger individual, the higher the magnitude of expenditure increase. This relationship is an indirect evidence for TPPS in mitigating MP.

### 3.2 Pension Participation, Livelihood Assets, and Multidimensional Poverty Mitigation

However, everyday expenditure $E_{ilt}^{C}$ can be not constant. Relaxing this assumption reveals that increased household consumption and investment does not imply poverty mitigation necessarily. It prompts this research to consider the relevance of this pathway to sustainable livelihoods with new structural economics. Expenditure behaviours such as wastefulness and gambling are unsustainable and cannot genuinely mitigate MP, while these behaviours are highly linked with non-contributory pension schemes and government transfers with meager market mechanisms (Haushofer & Shapiro, 2016). The TPPS is a combination of the facilitating state and effective markets; it raises market mechanisms of incentives and reduces welfare dependency, restraining the irrationality in household expenditures. The sustainable livelihoods framework offers a theoretical foundation and a testable channel (Natarajan et al., 2022). Chambers and Conway (1992) propose livelihoods as the means of living, built upon capabilities, assets, and activities. These sustainable livelihoods comprise two aspects: social and environmental. The social one implies the capacity to cope with and recover from pressures and shocks so that an individual could support future generations in the household. The environmental one means the ability to maintain and ameliorate local and global assets that underpin livelihoods, generating net livelihood benefits (Chambers & Conway, 1992).

On this sustainability categorising household capacities of livelihood activities, Scoones (1998) spotlights four capital types in the analytical framework: natural, human, financial, and social. Combining the studies by Sen (1999) as well Chambers and Conway (1992), the DFID (1999) furthers the sustainable livelihoods framework, subdividing financial capital into financial and physical. Considering the increasing value of psychological capital, this research incorporates psychological assets into our framework. At this point, our six capital assets enable our research to explore the channels and mechanisms of how TPPS impacts MP.

This research defines human capital as knowledge, labour capacity, and health status. Its accumulation could enhance productivity and increase income (Goldin, 2016), thereby breaking the vicious cycle of poverty. Next, social capital means how and where households improve economic returns (Knack & Keefer, 1997; Zhang et al., 2017) and access resources or career advice through social networks (Seibert et al., 2001). This research posits that the essence of social capital lies in fostering mutually beneficial cooperation through trust, norms, and networks. Then, physical capital includes the infrastructure and living assets necessary for sustaining livelihoods. As the most basic and direct form of assets, physical capital plays a crucial role in income and economic growth (Goode, 1959). Financial capital, such as cash, deposits, insurance, and stocks, implies the financial resources that people use to achieve their life goals. A mature financial system is vital for economic growth (King & Levine, 1993) and abundant financial capital broadens income sources to reduce poverty. Besides, according to DFID (1999), natural capital is the natural resources that generate resource flows and provide livelihood services. These range from intangible public goods such as the atmosphere and biodiversity to divisible assets like land for production. Natural capital is vital for households reliant on resource-based activities; for instance, land is intrinsically linked to poverty and poverty mitigation (Sharp, 2003). Last, psychological one stands for positive psychological attributes that influence personal development, including confidence, expectations of success, self-reliance, the courage to overcome difficulties, and optimism. A positive psychology could ameliorate work performance for more capital gains (Luthans et al., 2004; Li et al., 2020).

[Insert Table 4 here]

Referring to the analysed definitions of capitals and assets, this research employs the entropy method to calculate the scores of livelihood assets. Considering data availability and completeness, our livelihood analytical framework is constructed as shown in Table 4.

[Insert Figure 4 here]

To sum up, in a vulnerable context, a facilitating state leverages the endogeneity of individual characteristics, such as high saving rates, to establish and adjust the TPPS, which reduces uncertainties linked to ageing trends and external shocks. By combining state and market mechanisms, the TPPS offers households with a sense of security and stable future income. The stability subsequently substitutes precautionary savings and fosters consumption and investment in livelihood assets. The more pillars households participate in, the greater their expected income and the higher the likelihood of sustainable livelihoods development. Enhanced livelihood assets thus contribute to sustained improvements in the outcomes of education, health and living standards crucial for household MP mitigation (Fig. 4).

## 4. Research Design

### 4.1 Data

The sample used in our research is from the China Family Panel Studies. The CFPS is a biennial survey designed to be the Chinese equivalent of the US Panel Study of Income Dynamics. Our research selects the stage from 2012 to 2020 when the neoliberal approach within China's pension system reforms was modified with substantial state investment in social security that was relatively more urban-rural egalitarian (Liu & Sun, 2016). The Centre for Social Research at Peking University designs the CFPS questionnaire to cover the data from communities, households, individuals, adults, and children, capturing the dynamics of Chinese society, economy, education, and health.

Due to intergenerational impact caused by TPPS participation, this research integrates individual-level information to reflect the economic units at the household level. Outliers and missing values are carefully treated to preserve the integrity of the dataset. Efforts are made to recover lost information due to questionnaire navigation or from prior visits, maximising the retention of original data. The final dataset comprises 47,006 households across 31 provinces in China. It is crucial to note that this examination distinguishes between pension participation and collection. An individual is categorised as 'TPPS participant without collection' if confirms participation and is aged between 16 and 45. Additionally, for robustness checks, control variables of macroeconomic factors are introduced by matching with provincial-level data from the National Bureau of Statistics.

### 4.2 Variables

#### 4.2.1 Response Variables

Our research identifies the MP status as a dummy variable according to the measurement in Section 2. Drawing on the UNDP standard, the MP cutoff $k$ is 0.33 (Alkire & Foster, 2011). If an MP index measured exceeds the cutoff of 0.33, the variable is assigned a value of 1; otherwise, 0.

#### 4.2.2 Explanatory Variables

The explanatory variable is TPPS, measured from two perspectives. First, the number of pillars in which the household participates in pension system; second, whether a household is insured under a specific pillar of pension system. Based on the definition of the TPPS uncovered earlier, this research categorises the basic pension insurance, (old) rural pension insurance, new rural social pension insurance, and urban resident' pension insurance given by the CFPS as the first pillar. Enterprise annuity is the second pillar, and individual commercial insurance is categorised as the third one. A household is considered to have participated in a particular pillar if at least one member aged 16 to 45 is enrolled, with a binary assignment of 1 for participation and 0 for non-participation. Then the sum of the participation outcomes for the aforementioned three pillars is calculated, yielding the total number of pension pillars in which the household participates. This variable is an ordinal discrete variable ranging from 0 to 3.

#### 4.2.3 Control Variables

To enhance the robustness of the results, our research, with reference to available micro-level household data-based studies, controls variables at the head and general levels (Maitra & Ray, 2003; Alkire & Fang, 2019). The head of household manages its economic behaviours with a central role in financial decision-making, particularly in savings and participation in pension schemes. Thus, the characteristics of the head imply a direct impact on the decision-making of the household unit and its socioeconomic outcomes such as MP status.

At the general level, broader demographic and socioeconomic factors also operate. For example, the household size, the occupations of members, the gender composition, and the number of dependants introduce varying degrees of economic pressure and financial stability. The factors collectively affect how the household navigates financial choices of TPPS participation, offering additional layers of insight beyond the role of the head. In short, controlling variables at both the individual (head) and collective (general) levels enables our better identification of the factors driving household decision-making. The two-layered approach mitigates the risk of omitted variable bias and the potential for misattribution of explanatory variables, generating more robust estimations and a more comprehensive grasp of household TPPS participation.

Considering data availability, the specific controls are as follows. At the head level, controls are the household head's gender, health, age, and marital status. Involving the general level, controls extend to the total population of the household, the proportion of male members, the marriage rate among household members over 20 years old, the number of young and older dependants, whether any members are engaged in agriculture, whether any members are migrant workers, whether the household has any debt, and the household income-to-expense ratio.

### *4.3 Empirical Strategy*

The baseline identification strategy of this research exploits the participation status in the TPPS, inclusive of 47006 households across diverse regions in China. Given that the dependant variable, IFMP, is a dummy variable, employing a linear regression model would violate a key Gauss-Markov assumption, likely leading to heteroscedasticity in the model.

Consequently, following the research of Alkire and Fang (2019) and Liu et al. (2023), the research opts for a panel logit model to analyse the impact of TPPS on IFMP. However, unlike the random effects models employed in their studies, this work adopts a two-way fixed effects model (TWFE). The TWFE model can control for factors that are invariant over time and factors that are invariant across space, thereby mitigating the endogeneity issues that arise from omitted variables. With the advancement of econometrics, the use of fixed effects models has increasingly surpassed that of random effects models.

$$IFMP_{it} = \alpha_0 + \beta S(TPPS_{it}) + X_{it}\theta + \eta_i + \tau_t + \varepsilon_{it} \quad (10)$$

$$IFMP_{it} = \alpha_0 + \beta_1 TPPS\_1_{it} + \beta_2 TPPS\_2_{it} + \beta_3 TPPS\_3_{it} + X_{it}\theta + \eta_i + \tau_t + \varepsilon_{it} \quad (11)$$

Where $IFMP$ represents the MP status for household $i$, at time $t$. $TPPS\_1$、 $TPPS\_2$ and $TPPS\_3$ refers to the dummy variable for the state social insurance, enterprise annuity and individual commercial insurance participation of household $i$, in the pillar $n$, at time $t$. This binary indicator equals one if household engages in TPPS, and zero otherwise. The ordinal variable $S(TPPS)$ measures the degree of the TPPS participation from 0 (exclusion from TPPS) to 3 (full involvement in TPPS). Based on its control for the fixed effects of region (provincial $\eta$) and time (year $\tau$), TWFE model tackles endogeneity, isolates omitted variable bias, and therefore enhance the explainability of this research. This specification includes the control variables ($X$) from the aspects of households. $\varepsilon$ represents the cluster-robust standard errors.

## 5. Results

### 5.1 Descriptive Statistics

Table 5 reports the descriptive statistics for all variables in our model. The mean of the MP status is 0.413; over half of the households in the CFPS data have escaped from the MP status, but 41.3% still suffer from entrenched disparities in certain dimensions. The mean household participation in the TPPS accounts for 0.634, and that in the first pillar is 0.583, indicating a relatively high coverage of state social insurance. Although this mean might underestimate the coverage due to CFPS data availability, it at least implies the failure of China's public pension schemes to achieve universal coverage.

[Insert Table 5 here]

Moreover, the mean values for the second (enterprise annuity) and third (individual commercial insurance) pillars are respectively 0.036 and 0.032, which clarified the significant disparities in household participation. The mean and median values of other variables portray the characteristics of our household samples.

*5.2 Baseline Regressions*

Table 6 reports the baseline regressions. The response variable in the Table 6 is $IFMP$. The explanatory variable in Columns (1) is $S(\ TPPS\ )$, and those in (2), (3), (4) and (5) are $TPPS\_1$, $TPPS\_2$, and $TPPS\_3$ respectively and all. Columns (2), (3), and (4) examine the influence of state social insurance, enterprise annuity, and individual commercial insurance on the household MP status, assuming there is no co-influence among the pillars; Column (5) incorporates all three pillars into the model. The coefficients in (1), (2), (3), (4) and (5) are significantly negative, indicating that the more multiple a household engages in the TPPS, the less probability it suffers from the MP status.

Observing the results in Columns (2), (3), (4) and (5), the influence of state social insurance is weaker than that of enterprise annuity and commercial insurance. Regarding our available knowledge articulated in Sections 1, 2, and 3, changes in confidence, uncertainty, and risk are important determinants. The enterprise annuity and commercial insurance, as market-oriented pension schemes, offer greater subjectivity in the engagement compared to state public pension ones. With higher expectations of future payments, participants believe the second and third pillars as substantial old-age security. This belief increases current engagement in anti-risk and -uncertainty actions to improve the household MP status. Therefore, enterprise annuity and commercial insurance ought to realise a higher degree of coverage compared to social insurance with weaker capabilities to mobilise the subjective expectations and objective actions of participants.

[Insert Table 6 here]

The findings also deliver evidence for the social policy inspiration by new structural economics. The state pension system is mandatory in accordance with national policy, while participation in enterprise annuity and commercial insurance is determined by individuals, employer enterprises, and insurance companies. The role of state social insurance in MP mitigation underscores the significance of state intervention; however, the results of the first pillar reflect that its extensive coverage but limited support hinder the influence of the first pillar. The impacts of enterprise annuity and commercial insurance manifest the must of facilitating the effective markets. Nonetheless, according to our descriptive statistics, the coverage of TPPS_2 and TPPS_3 is limited, which requires the facilitating state to consider the sustainability of pension system and the efficiency of the markets. As the preliminary findings drawn from the baseline regressions may be subject to skepticism, the following sections further check and test the endogeneity and general robustness to verify the reliability of our conclusions.

*5.3 Endogeneity Checks*

Certain concerns might weaken the argument rooted in baseline identification from (1) to (4). First, MP may in turn affect the extent of participation in the TPPS. Given the voluntary nature of the second and third pillars in the TPPS, participation could be an endogenous household decision. Besides, the poverty status of a household plays a role: a wealthier household is often associated with higher levels of education and health with greater financial literacy and a better understanding of the expected returns from various pension schemes. Higher levels of health also enable such households to realise greater pension benefits, thus encouraging more active participation in the different pension pillars. However, low levels of pension benefits may not appeal to higher-income groups, and affluent households might lack interest in the three pillars altogether.

Second, both the TPPS participation and household-level poverty seem subject to unobservable elements, and available China's individual-level longitudinal surveys exclude relevant data, as the CFPS could be unable to objectively measure households' sunk cost, mental accounting, hot hand fallacy, or availability heuristic inherent in economic behaviours in pension studies. The sunk cost and hot hand fallacy further amplify this sensitivity and aversion before households attempt to more comprehensively engage in TPPS. Thanks to intergenerational experience and existing dependency on the state, the households in China often believe previous success of the state social insurance to endure indefinitely. Such availability heuristic misguides people to narrowly focus on previous successful dependency on government for generations, drifting mass perception and behaviours away from TPPS.

*5.3.1 Instrumental Variables*

The concerns direct our research to econometric endogeneity. To strengthen the explainability of our TWFE model, this check deals with endogeneity by using the data of TPPS pillars per household at the county level as the instrumental variable (IV). It employs the shift-share method to satisfy the relevance and exogeneity conditions. Specifically, the product of the per capita number of TPPS participating within a district where the sample household's address is located and the national per capita change rate of TPPS participation from year $t−1$ to t is used as the IV for the participation quantity. Additionally, the product of the participation rate of each specific pillar in a district where the household's address is located and that change rate of participation in the specific pillar from $t−1$ to t is also used for separate pillars.

The constructed IVs are highly correlated with household TPPS participation, reflecting the cognition and behavior of households in specific regions, and are not directly causally related to household MP status. This construction method draws on the latest theoretical achievements of Bartik IV (Goldsmith-Pinkham et al., 2020; De Chaisemartin & Lei, 2023). Goldsmith-Pinkham et al. (2020) note that while it is ideal to determine the exogeneity of both the share and the shock using the Bartik IV, such a coincidence may be difficult to achieve in reality. They suggest that satisfying one of the assumptions is a more practical strategy; although the mean participation at the district level may not meet the exogeneity condition of the share, the national participation rate change has exogeneity, and the share and exogenous shock are a linear combination—which can yield unbiased and consistent estimates. Although using a time-varying share deviates from the original Bartik IV, it has been widely adopted and proven to be equally valid as the conclusions drawn from a time-invariant share when a shock occurs between $t−1$ and t (De Chaisemartin & Lei, 2023).

*5.3.2 Conditional Mixed Process*

Here the TPPS is an ordinal discrete variable, and TPPS_1, TPPS_2, and TPPS_3 are dummy ones. Given the variables, the IV-probit two-step model is not applicable. Therefore, this research employs the Conditional Mixed Process (CMP) estimation proposed by Roodman (2011) to address the endogeneity. This estimation uses maximum likelihood estimation (MLE) and treats the equations as a whole. Due to the flexibility of MLE, the CMP accommodates coefficient constraints as well as various types of weights and clustered robust standard errors at different levels. Initially, the first-stage regressions examine the relationship between the IVs and the endogenous variable, with the response variables being the number of pillars in which a household participates in and whether in the specific pillars. The explanatory variables include the corresponding IVs, controls, and TWFE. Subsequently, second-stage regressions employ IV approach to re-estimate the impact of household TPPS participation on MP status.

[Insert Table 7 here]

Table 7 presents the IV-included estimated effects of TPPS participation on the status of MP. The endogenous variables and their corresponding IVs regression results are both significantly positive at the 1% level, indicating a strong correlation between the instrumental and endogenous variables. Additionally, Table 7 shows that the CMP statistic, atanhrho, is significant at the 5% level, suggesting that TPPS, TPPS_1, TPPS_2, and TPPS_3 are endogenous variables. Hence, the selection of IVs is effective. Compared with the estimated results of the logit model, after addressing the endogeneity of the model, the estimated values (marginal effects) of TPPS, TPPS_1, TPPS_2, and TPPS_3 are significantly negative, but the absolute values of the estimated influence increase. The logit model underestimated the influence of TPPS, TPPS_1, TPPS_2, and TPPS_3 on IFMP, and the CMP results are more consistent and effective. The improvement effect of TPPS_1 on MP status is significantly weaker than that of TPPS_2 and TPPS_3, consistent with the baseline regressions.

*5.4 Robustness Tests*

The robustness of the baseline regressions forms a key concern. Our robustness checks are based on the model that has accounted for endogeneity. In our endogeneity checks, IV and CMP approaches are deployed to enhance the reliability of the findings. Therefore, the robustness tests here winsorise sample size, change the timeframe, adjust the level of clustering for standard errors, substitute response variables, substitute explanatory variables, and add control variables at the macroeconomic level to check the explainability of the regression findings. Particularly, concerning arise regarding the selection into specific pillars being influenced by household characteristics, this research employs propensity score matching to mitigate these concerns.

After the robustness tests in seven conditions, our results exhibit no significant change, indicating a certain robustness degree of our empirical strategy. To strengthen scientific communication efficiency, the results are reported in Appendices.

### 5.5 Heterogeneity Trials

Our research explores the differential effects of TPPS participation on MP across urban and rural settings, necessitating a split-sample examination to capture the variations. To account for endogeneity, the heterogeneity trials employ both IV and CMP approaches. Columns (1) and (2) of Table 8 examine the effects of TPPS participation degree in rural and urban areas, while Columns (3) and (4) focus on the specific variations for each TPPS pillar across these areas. The consistent results reveal that higher TPPS participation brings about a lower MP index, a trend verified across both urban and rural households. The first and second pillars of TPPS mitigate poverty in both, though the third pillar fails to do so in rural areas. Our Chow test further highlights intergroup coefficient variations, with significant values in Columns (1) and (2), indicating a stronger impact of TPPS participation on MP mitigation in urban households compared to rural ones. In Columns (3) and (4), only the coefficient for the first pillar is significant, suggesting its efficacy in urban areas rather, or at least more, than in rural, while no significant variations are tested for the second and third pillars.

[Insert Table 8 here]

One explanation is the structural disparities within the pension system. As noted in the background, urban pension schemes benefited from earlier facilitative intervention by the state. The first pillar's partial funding scheme, alongside various tax incentives for the second and third pillars, harmoniously integrates state and market roles. By contrast, rural pension structures are comparatively nascent, with the first pillar only stabilising post-2014 and the second and third pillars very underdeveloped. Although the recent *Rural Revitalisation Strategy* introduces more enterprises in rural areas, bolstering rural pensions, the insignificant Chow test results suggest that the second pillar in rural areas is not yet as impactful as in urban regions. The insignificance of the third pillar reflects market dynamics, wherein the commercial insurers, acting rationally, often avoid rural markets, which triggers supply-side disparities and the subsequent underperformance of the third pillar. Ultimately, a two-pillar system without complete market-state alignment limits its overall efficacy.

## 6. Mechanisms

### *6.1 Livelihood Assets*

This section uncovers the specific mechanisms underlying the effects of TPPS on MP status. Livelihood assets refer to the concept, sustainable livelihoods, guiding our anti-MP research. This concept works within the sustainable livelihoods framework, through which the research embeds the TPPS initiative in the context of development (Scoones, 1998; DFID, 1999; Sen, 1999). Livelihoods, together with the capabilities and assets necessary to sustain livelihoods, are sustainable when capable of being maintained and recovered from shocks without undermining the future. Livelihood assets position social, human, and financial capital at the centre of the framework, in favour of a tacit recognition of capitalism (Natarajan et al., 2022). Whilst the available literature furthers the exploration of relational power and environmental relations, this research integrates the adjustments and changes within the framework of livelihood assets, not to demonstrate any ideological attitude towards capitalism but rather to characterise the sustainable livelihoods framework through the medium of these assets: people-centred, multidimensional, dynamic, and cooperative. Our empirical strategy contains relevant assets of human, social, natural, physical, financial, and psychological ones.

To examine the TPPS participation increases the probability of livelihood asset development—investment and consumption to acquire or improve multiple dimensions

assets—our strategy counts on microeconomic data, calculating the scores ( *ZScore* ) for each household's livelihood assets per year. Estimating how TPPS affects the *ZScore* , our TWFE model is engaged:

$$ZScore_{it} = \alpha_0 + \beta S(TPPS_{it}) + X_{it}\theta + \eta_i + \tau_t + \varepsilon_{it} \quad (12)$$

$$ZScore_{it} = \alpha_0 + \beta_1 TPPS\_1_{it} + \beta_2 TPPS\_2_{it} + \beta_3 TPPS\_3_{it} + X_{it}\theta + \eta_i + \tau_t + \varepsilon_{it} \quad (13)$$

As shown in Table 9, the Column (1) indicates that the more pillars a household participate in, the more developed their livelihood assets become. Columns (3) and (4) examine the effects of the participation in enterprise annuity and commercial insurance to livelihood asset development. The promising results not only identify the TPPS-MP bridge but also demonstrate the sustainable poverty mitigation outcomes caused by TPPS. The results act as empirical evidence for the effects of TPPS as an anti-MP policy in China until 2020. Different from economic growth and cash transfer, the improved livelihood assets enhance the resources available to previously vulnerable communities, strengthening their resilience capable of recovering from exogenous shocks such as global health crises. Well addressing the vulnerability risk and resilience capability offers more opportunities to build sustainable livelihood assets for all of the generations within given households.

[Insert Table 9 here]

However, the results of state social insurance fail to be significant, which remains consistent with our theoretical reasoning. Due to the extensive coverage mandated by government responsibility, the state pension system promotes its coverage at the expense of future returns. When individuals anticipate fewer returns, they would shift money into precautionary savings, diminishing investment and consumption for developing livelihood assets. In other words, the security and sustainability of state social insurance might be distrusted. Accordingly, it is insignificant to the holistic livelihood asset mechanism, which calls for further dissection in our next section. Nevertheless, household participation in market-driven pension schemes such as TPPS_2 and TPPS_3 makes up for the deficiency of security based on more sources and resources involved in the markets, delivering greater incentives for households to invest in and consume for livelihood assets to sustain resilient against uncertainties and risks.

From the sustainable livelihoods framework, the importance of combining a facilitating state with the effective markets is evident. Social policy was considered within the purview of government responsibility, but our evidence indicates that the involvement of market actors is indispensable. Compared to state intervention, the power of markets is more capable of driving households to make investment decisions and adjust consumption choices, shifting more expenditures into sustainable livelihoods development. In this regard, the TPPS participation is involved in the decisions, mitigating the risk of falling into MP status. Therefore, the facilitating state ought to introduce market actors in social policy praxes to effectively enhance the security and sustainability of the pension system.

### 6.2 Heterogeneous Assets

Targeted at the holistic mechanism through which TPPS mitigates MP, this section is curious about detailed mechanisms of the asset-based improvement on each component. The scores ( $YScore$ ) for heterogeneous assets and capital of human (HScore), social (SScore), natural (NScore), physical (PHScore), financial (FScore), and also psychological (PSScore) ones are examined separately.

$$YScore_{it} = \alpha_0 + \beta S(TPPS_{it}) + X_{it}\theta + \eta_i + \tau_t + \varepsilon_{it} \quad (14)$$

$$YScore_{it} = \alpha_0 + \beta_1 TPPS\_1_{it} + \beta_2 TPPS\_2_{it} + \beta_3 TPPS\_3_{it} + X_{it}\theta + \eta_i + \tau_t + \varepsilon_{it} \quad (15)$$

Table 10 indicates that the more comprehensive a household's participation in TPPS, the more developed their livelihood assets are in human, natural, financial, and psychological capital. Since life cycle and permanent income security improve, Chinese households are more inclined to prioritise investment in and consumption for human, natural, financial, and psychological aspects. Human capital includes compulsory education and vocational training and healthcare, which are crucial for enhancing the education, health, and decent work levels within a household. Natural capital involves the environment, land, and clean water. To be specific, consuming clean water, renting arable land, and protecting our natural environment are instrumental in seeking a higher living standard. Financial capital comprises savings, bonds, stocks, and so forth offered by financial institutions, serving as diverse avenues to broaden household income sources. Besides, development-oriented consumption choices elevate psychological capital, increasing wellbeing and confidence in the present and future.

[Insert Table 10 here]

The livelihood assets and capital statistically signify that, under the catalysis of TPPS, Chinese households become more rational in poverty mitigation actions. The investment in human, natural, and financial capital is meaningful, as the low given levels of capital are a major challenge in the South states. Those in livelihood assets not only mitigate the household MP status but also strengthen their capabilities to sustain resilient against external risks and uncertainties, enhancing the security and sustainability of MP mitigation.

[Insert Table 11 here]

Table 11 concludes that the changes caused by enterprise annuity and commercial insurance participation are more significant. The two market-focused schemes almost impact every single capital development. The results of social insurance are significant only in the regressions for human, natural, and financial capital. Under state's priority of universal coverage, its coverage rate in China is relatively high. However, higher coverage and China's population trigger a decline in pension money. Enterprise annuity and commercial insurance augment additional security for the later years of retirement; this security answers why the effective markets gain increasing importance in livelihood assets. The coefficients and their significance reveal that, compared to TPPS_1, TPPS_2, and TPPS_3 have a more evident influence on stimulating investment in and consumption for livelihood asset development. In sustainable MP mitigation, the two market-focused schemes play an indispensable role.

In summary, although the effects of TPPS spark a broader discussion, promoting social participation in TPPS and improving social policy measures are not exclusive. Certain indicators might not always align with simplified linear causality; hence, policymakers are advised to engage a pragmatic facilitating state perspective, taking more selective measures to secure and develop people's sustainable livelihoods and mitigate MP—which is more beneficial than being entangled with technical or ideological debates.

## 7. Discussion & Conclusions

Our research highlights the critical role of a balanced three-pillar system of pension provision in MP mitigation coordinated by the state and markets. First, drawing on the sustainable livelihoods framework and new structural economics, comprehensive TPPS participation mitigates MP risks by fostering household expenditures on human, natural, financial, and psychological capital. Although state social insurance, enterprise annuity, and commercial insurance all contribute to MP mitigation, the market-oriented pillars—enterprise annuity and commercial insurance—prove an evident and robust impact, correlating to consumption and investment in livelihood assets that strengthen household resilience against uncertainties and shocks. Specifically, the stronger impact of social insurance in urban areas constitutes the urban-rural heterogeneous effects of TPPS participation when commercial insurance fails to do so in rural contexts. This heterogeneity seems due to the structural disparities within the pension system: urban pension schemes benefit from more state intervention; by contrast, rural ones are relatively nascent and underdeveloped.

Second, the findings align with new structural economics, where state intervention complements market mechanisms to settle structural gaps. The evolution of China's pension system from state-dominated solo to multi-player ceilidh reflects a proactive, selective approach that consolidates the sustainability and reliability of social security, ensuring a stable income for retirees to satisfy their demands. Regarding the pension system reform and broader policy implications, it is imperative to enhance the dynamism and coordination akin to the ceilidh. China's experience offers a valuable reference for other South states when encountering similar challenges in building a resilient social policy tango.

However, caution must temper the increasing financialisation of welfare. Although market-oriented pillars are essential, the hallmark of the neoliberal approach risks shrewdly shifting pension responsibility to individuals, embedding market risks in social welfare and downplaying inequalities. Thus, our research advocates a ceilidh approach where the state aligns market mechanisms with sustainable development goals—pension reforms must prioritise long-term stability over short-term returns. Additionally, policymakers frequently encounter the dilemma of implementing necessary yet unpopular reforms. To settle the dilemma, options such as closed-door negotiations and automatic stabilisers can consolidate the delicate coordination between fiscal stability and the social contract that underpins public support for pension reforms. Nevertheless, the transparency and inclusivity of reforms raise criticism. With new structural economics, a facilitating state must navigate policymaking with the demographic realities to coordinate that reforms are not only technically solid but socio-politically sustainable—ultimately fostering public trust and enhancing the efficacy of social policy in mitigating poverty and inequality.

Although our research shares valuable insights into the role of the TPPS in mitigating MP, several limitations are also acknowledged. 1) The research design leaves out behavioural factors, such as financial literacy, risk tolerance, and attitudes towards savings and retirement. These factors strongly affect how households participate in the TPPS and thus the system's efficacy to mitigate poverty. Incorporating behavioural economics can reveal the intricate processes of household decision-making, enabling the development of more targeted policy measures following individual preferences and behaviours. 2) The potential impact of external shocks, such as economic crises or pandemics, is vital. Future research is warranted to integrate those variables better to understand the resilience of the TPPS under stressful conditions. Focused on the shocks, comparative studies across cultural and institutional contexts can reveal the universality and particularity of various pension systems. 3) Further exploring the interplay between different livelihood assets and their combined impact on MP can uncover relationships among and the collective role of various capital dimensions such as human, natural, and financial ones. It will facilitate a holistic understanding of how pension systems improve the overall wellbeing of households and inform anti-MP policymaking in a multifaceted way.

In conclusion, the TPPS acts as a vital instrument in mitigating MP, with its impact extending to various dimensions of human wellbeing. The findings reinforce the necessity of a holistic approach to social policy, where the facilitating state aligns market supplies with the people's demands. This research contributes to the ongoing dialogue on the role of social policy measures and the pension system in fostering nuanced and actionable inclusive development, offering theoretical and practical insights for sustainable poverty mitigation in the South. It is called to scrutinise social security in the context of poverty mitigation, particularly within socioeconomic and demographic structures.

**Classification Codes**

JEL D1 Household Finance, I3 Government Welfare Policy, O1 Economic Development

**Declaration of Interest Statement**

The authors declare no conflicts of interest.

**Acknowledgements**

The authors wish to express supreme appreciation to Xianming Fang, Lingguo Cheng, and Yongfan Li at the Nanjing University Department of Finance for their research assistance and advice, and several referees for valuable comments on and for improving our research. Many thanks to the Peking University Centre for Social Research for data collection. This work was supported by the National Social Science Fund of China Programme (Grant ID Number: 21BGL291) and Major Programme (Grant ID Number: 24&ZD092). The responsibility for the content and any errors, naturally, remains exclusively with the authors.

**Tables**

**Table 1.** *MP Dimensions and Indicators*

| Dimension | Indicator | Explanation | Weight |
|---|---|---|---|
| Education | Average Years of Education | Whether the average years of education for members over 16 was less than 9 | 1/6 |
| | School Enrollment | Proportion of children (aged 6–16) who dropped out of school | 1/6 |
| Health | Body Mass Index (BMI) | Whether more than 50% of the adults in the household had a BMI not within the range of 18.5–25 | 1/12 |
| | Hospitalisation | Whether someone in the household was hospitalised in the past year | 1/12 |
| | Self-Assessed Health | Whether someone in the household considered themselves to be in very poor health | 1/12 |
| | Health Insurance | Whether someone in the household was not covered by health insurance | 1/12 |
| Living Standards | Cooking Fuel | Whether the household had no access to clean fuel for cooking | 1/15 |
| | Water Access | Whether the household had no access to tap water, bottled water, pure water, or filtered water | 1/15 |
| | Housing Space | Whether the per capita housing area of the household was less than 12 square metres | 1/15 |
| | Electricity Access | Whether the household had no access to electricity | 1/15 |
| | Durable Goods Value | Whether the total value of durable goods in the household was less than 1000 CNY | 1/15 |

**Table 2.** *MP Incidence*

| Heterogeneity | Category | Year | k=0.33 |
|---|---|---|---|
| | | 2012 | 0.494 |
| | | 2014 | 0.431 |
| | H | 2016 | 0.433 |
| | | 2018 | 0.391 |
| | | 2020 | 0.297 |
| | | 2012 | 0.437 |
| | | 2014 | 0.432 |
| All Households | A | 2016 | 0.427 |
| | | 2018 | 0.422 |
| | | 2020 | 0.418 |
| | | 2012 | 0.216 |
| | | 2014 | 0.186 |
| | $M_0$ | 2016 | 0.185 |
| | | 2018 | 0.165 |
| | | 2020 | 0.124 |

| Heterogeneity | Category | Year | k=0.33 | Heterogeneity | Category | Year | k=0.33 |
|---|---|---|---|---|---|---|---|
| | | 2012 | 0.598 | | | 2012 | 0.375 |
| | | 2014 | 0.543 | | | 2014 | 0.318 |
| | H | 2016 | 0.569 | | H | 2016 | 0.308 |
| | | 2018 | 0.519 | | | 2018 | 0.280 |
| | | 2020 | 0.427 | | | 2020 | 0.203 |
| | | 2012 | 0.445 | | | 2012 | 0.419 |
| | | 2014 | 0.440 | | | 2014 | 0.412 |
| Rural Households | A | 2016 | 0.439 | Urban Households | A | 2016 | 0.409 |
| | | 2018 | 0.434 | | | 2018 | 0.404 |
| | | 2020 | 0.429 | | | 2020 | 0.399 |
| | | 2012 | 0.266 | | | 2012 | 0.157 |
| | | 2014 | 0.239 | | | 2014 | 0.131 |
| | $M_0$ | 2016 | 0.250 | | $M_0$ | 2016 | 0.126 |
| | | 2018 | 0.225 | | | 2018 | 0.113 |
| | | 2020 | 0.183 | | | 2020 | 0.081 |

**Table 3.** *MP Dissection*

| Year | Dimension | All Households | | Rural Households | | Urban Households | |
|---|---|---|---|---|---|---|---|
| | | k=0.33 | k=0.66 | k=0.33 | k=0.66 | k=0.33 | k=0.66 |
| 2012 | Education | 0.385 | 0.340 | 0.383 | 0.343 | 0.390 | 0.333 |
| | Health | 0.401 | 0.349 | 0.371 | 0.330 | 0.460 | 0.392 |
| | Living Standards | 0.214 | 0.312 | 0.264 | 0.327 | 0.151 | 0.275 |
| 2014 | Education | 0.387 | 0.346 | 0.382 | 0.341 | 0.395 | 0.341 |
| | Health | 0.409 | 0.398 | 0.384 | 0.390 | 0.457 | 0.427 |
| | Living Standards | 0.204 | 0.256 | 0.234 | 0.263 | 0.148 | 0.232 |
| 2016 | Education | 0.387 | 0.316 | 0.381 | 0.316 | 0.399 | 0.317 |
| | Health | 0.419 | 0.404 | 0.394 | 0.405 | 0.467 | 0.397 |
| | Living Standards | 0.194 | 0.280 | 0.225 | 0.280 | 0.134 | 0.286 |
| 2018 | Education | 0.391 | 0.280 | 0.380 | 0.297 | 0.409 | 0.240 |
| | Health | 0.434 | 0.421 | 0.410 | 0.413 | 0.479 | 0.440 |
| | Living Standards | 0.176 | 0.299 | 0.210 | 0.290 | 0.112 | 0.320 |
| 2020 | Education | 0.394 | 0.321 | 0.386 | 0.333 | 0.408 | 0.271 |
| | Health | 0.436 | 0.413 | 0.413 | 0.410 | 0.476 | 0.435 |
| | Living Standards | 0.170 | 0.266 | 0.200 | 0.256 | 0.115 | 0.348 |

**Table 4.** *Sustainable Livelihood Assets Dimensions and Indicators*

| Type | Indicator | Explanation | Mean | SD |
|------|-----------|-------------|------|-----|
| Human Capital (H) | Years of Education (H1) | Average years of education for members over 16 | 7.731 | 4.126 |
| | Health Investment (H2) | Expenditures on healthcare | 301.3 | 1983 |
| Social Capital (S) | Gift Income (S1) | Household incomes from gifts | 1567 | 8201 |
| | Gift Expenditure (S2) | Household expenditures on gifts | 2959 | 5580 |
| | Social Status (S3) | Mean self-assessed social status of members over 16 | 2.903 | 0.867 |
| Physical Capital (PH) | Durable Goods Expenditure (PH1) | Household expenditures on durable goods | 1562 | 18184 |
| | Daily Necessities Expenditure (PH2) | Household expenditures on daily necessities | 73.62 | 99.59 |
| Financial Capital (F) | Deposits (F1) | Household deposits in financial institutions | 39795 | 120392 |
| | Financial Products (F2) | Household investments in stocks, funds, government bonds, trust products, foreign exchange products, and other internet financial products | 0.0660 | 0.248 |
| Natural Capital (N) | Land Ownership (N1) | Household ownership of at least one type of land: arable land, forest land, pasture, or pond | 0.623 | 0.485 |
| | Land Rental Expenditure (N2) | Rent paid for land used in production | 268.5 | 4129 |
| Psychological Capital (PS) | Life Satisfaction (PS1) | Mean life satisfaction of members over 16 | 3.708 | 0.875 |
| | Future Confidence (PS2) | Mean future confidence level of members over 16 | 3.938 | 0.856 |

**Table 5.** *Descriptive Statistics*

| Variable | N | Mean | P50 | SD | Min | Max |
|---|---|---|---|---|---|---|
| IFMP | 47006 | 0.413 | 0 | 0.492 | 0 | 1 |
| TPPS | 47006 | 0.634 | 1 | 0.546 | 0 | 3 |
| TPPS 1 | 47006 | 0.583 | 1 | 0.493 | 0 | 1 |
| TPPS 2 | 47006 | 0.036 | 0 | 0.187 | 0 | 1 |
| TPPS 3 | 47006 | 0.032 | 0 | 0.177 | 0 | 1 |
| householdsize | 47006 | 3.679 | 3 | 1.822 | 1 | 17 |
| marryrate | 47006 | 0.754 | 1 | 0.319 | 0 | 1 |
| sumjuvenile | 47006 | 0.784 | 1 | 0.960 | 0 | 8 |
| sumolder | 47006 | 0.486 | 0 | 0.743 | 0 | 4 |
| malerate | 47006 | 0.511 | 0.5 | 0.198 | 0 | 1 |
| ifpeasant | 47006 | 0.489 | 0 | 0.500 | 0 | 1 |
| ifmigrantworker | 47006 | 0.341 | 0 | 0.474 | 0 | 1 |
| lnearnrate | 47006 | 0.950 | 0.847 | 0.537 | 0 | 11.18 |
| ifdebt | 47006 | 0.146 | 0 | 0.353 | 0 | 1 |
| headage | 47006 | 54.98 | 54 | 17.52 | 14 | 120 |
| headgender | 47006 | 0.594 | 1 | 0.491 | 0 | 1 |
| headmarry | 47006 | 0.852 | 1 | 0.355 | 0 | 1 |
| headhealth | 47006 | 0.172 | 0 | 0.377 | 0 | 1 |
| PSScore | 47006 | 0.887 | 0.877 | 0.255 | 0.002 | 3.247 |
| SScore | 47006 | 0.037 | 0.03 | 0.037 | 0.001 | 0.996 |
| HScore | 47006 | 0.022 | 0.02 | 0.022 | 0 | 0.984 |
| FScore | 47006 | 0.093 | 0.039 | 0.182 | 0.002 | 2.439 |
| PHScore | 47006 | 0.047 | 0.037 | 0.047 | 0.002 | 1.738 |
| NScore | 47006 | 0.158 | 0.125 | 0.178 | 0.002 | 2.439 |
| ZScore | 47006 | 1.243 | 1.121 | 0.611 | 0.01 | 10.78 |

**Table 6.** *TPPS Participation on MP Status*

| VARIABLES | (1) | (2) | (3) | (4) | (5) |
|---|---|---|---|---|---|
| TPPS | -0.231*** | | | | |
| | (0.030) | | | | |
| TPPS_1 | | -0.145*** | | | -0.141*** |
| | | (0.033) | | | (0.033) |
| TPPS_2 | | | -0.676*** | | -0.655*** |
| | | | (0.086) | | (0.086) |
| TPPS_3 | | | | -0.548*** | -0.509*** |
| | | | | (0.085) | (0.085) |
| Constant | -4.572*** | -4.640*** | -4.664*** | -4.676*** | -4.569*** |
| | (0.271) | (0.272) | (0.271) | (0.272) | (0.271) |
| Controls | Y | Y | Y | Y | Y |
| TWFE | Y | Y | Y | Y | Y |
| Observations | 47,006 | 47,006 | 47,006 | 47,006 | 47,006 |

Notes: *** p < 0.01, ** p < 0.05, * p< 0.1. The columns reports the estimates of Expr. (1) and (2). All regressions apply TWFE model. Robust standard errors are clustered at the household level. Hereinafter the same.

**Table 7.** *IV-CMP Checks*

| VARIABLES | (1) TPPS | (2) IFMP | (3) TPPS_1 | (4) TPPS_2 | (5) TPPS_3 | (6) IFMP |
|---|---|---|---|---|---|---|
| TPPS | | -0.610*** | | | | |
| | | (0.062) | | | | |
| BartikTPPS | 0.787*** | | | | | |
| | (0.017) | | | | | |
| TPPS_1 | | | | | | -0.242*** |
| | | | | | | (0.065) |
| TPPS_2 | | | | | | -3.350*** |
| | | | | | | (0.226) |
| TPPS_3 | | | | | | -0.506* |
| | | | | | | (0.276) |
| BartikTPPS_1 | | | 0.783*** | | | |
| | | | (0.016) | | | |
| BartikTPPS_2 | | | | 0.530*** | | |
| | | | | (0.032) | | |
| BartikTPPS_3 | | | | | 1.390*** | |
| | | | | | (0.082) | |
| Controls | Y | Y | Y | Y | Y | Y |
| TWFE | Y | Y | Y | Y | Y | Y |
| Observations | 39,020 | 39,020 | 39,020 | 39,020 | 39,020 | 39,020 |
| atanhrho | 0.229*** | - | 0.080*** | 0.674*** | 0.097** | - |

**Table 8.** *TPPS Participation on MP Status of Heterogeneous Households*

| VARIABLES | (1) Rural | (2) Urban | (3) Rural | (4) Urban |
|---|---|---|---|---|
| TPPS | -0.350*** | -0.677*** | | |
| | (0.097) | (0.089) | | |
| TPPS_1 | | | -0.184* | -0.319*** |
| | | | (0.109) | (0.093) |
| TPPS_2 | | | -3.373*** | -2.740*** |
| | | | (0.641) | (0.270) |
| TPPS_3 | | | 0.072 | -0.536* |
| | | | (0.522) | (0.314) |
| Chow Test | | -0.163*** | 0.143*** | -0.025 | -0.003 |
| Controls | Y | Y | Y | Y |
| TWFE | Y | Y | Y | Y |
| Observations | 18,704 | 18,704 | 19,621 | 19,621 |

Notes: The results of the Chow test are based on the estimated values with interaction terms. The results based on Fisher's Permutation test remain consistent.

**Table 9.** *TPPS Participation on Livelihood Assets*

| VARIABLES | (1) ZScore | (2) ZScore | (3) ZScore | (4) ZScore | (5) ZScore |
|---|---|---|---|---|---|
| TPPS | 0.0239*** | | | | |
| | (0.0061) | | | | |
| TPPS_1 | | 0.0090 | | | 0.0094 |
| | | (0.0062) | | | (0.0062) |
| TPPS_2 | | | 0.0765*** | | 0.0744*** |
| | | | (0.0188) | | (0.0188) |
| TPPS_3 | | | | 0.0649*** | 0.0615*** |
| | | | | (0.0217) | (0.0217) |
| Constant | 1.2557*** | 1.2631*** | 1.2649*** | 1.2665*** | 1.2604*** |
| | (0.0327) | (0.0327) | (0.0326) | (0.0326) | (0.0327) |
| Controls | Y | Y | Y | Y | Y |
| Observations | 43,514 | 43,514 | 43,514 | 43,514 | 43,514 |
| R-Squared | 0.638 | 0.638 | 0.639 | 0.638 | 0.639 |

**Table 10.** *TPPS Participation on Heterogeneous Assets*

| VARIABLES | (1)<br>HScore | (2)<br>SScore | (3)<br>NScore | (4)<br>PHScore | (5)<br>FScore | (6)<br>PSScore |
|---|---|---|---|---|---|---|
| TPPS | 0.0010*** | 0.0005 | 0.0089*** | 0.0009 | 0.0078*** | 0.0048* |
|  | (0.0002) | (0.0004) | (0.0018) | (0.0006) | (0.0018) | (0.0026) |
| Constant | 0.0242*** | 0.0452*** | 0.1524*** | 0.0579*** | 0.1038*** | 0.8722*** |
|  | (0.0009) | (0.0025) | (0.0095) | (0.0034) | (0.0092) | (0.0140) |
| Controls | Y | Y | Y | Y | Y | Y |
| TWFE | Y | Y | Y | Y | Y | Y |
| Observations | 43,514 | 43,514 | 43,514 | 43,514 | 43,514 | 43,514 |
| R-Squared | 0.713 | 0.522 | 0.618 | 0.510 | 0.669 | 0.601 |

**Table 11.** *Three Pillars Participation on Heterogeneous Assets*

| VARIABLES | (1)<br>HScore | (2)<br>SScore | (3)<br>NScore | (4)<br>PHScore | (5)<br>FScore | (6)<br>PSScore |
|---|---|---|---|---|---|---|
| TPPS_1 | 0.0005*** | -0.0002 | 0.0052*** | -0.0002 | 0.0041** | 0.0001 |
|  | (0.0002) | (0.0004) | (0.0018) | (0.0006) | (0.0018) | (0.0028) |
| TPPS_2 | 0.0019*** | 0.0029** | 0.0207*** | 0.0042** | 0.0201*** | 0.0247*** |
|  | (0.0007) | (0.0012) | (0.0058) | (0.0016) | (0.0057) | (0.0070) |
| TPPS_3 | 0.0019** | 0.0017 | 0.0198*** | 0.0044** | 0.0180*** | 0.0157* |
|  | (0.0008) | (0.0013) | (0.0067) | (0.0019) | (0.0065) | (0.0080) |
| Constant | 0.0243*** | 0.0455*** | 0.1536*** | 0.0582*** | 0.1050*** | 0.8737*** |
|  | (0.0009) | (0.0025) | (0.0095) | (0.0034) | (0.0091) | (0.0140) |
| Controls | Y | Y | Y | Y | Y | Y |
| TWFE | Y | Y | Y | Y | Y | Y |
| Observations | 43,514 | 43,514 | 43,514 | 43,514 | 43,514 | 43,514 |
| R-Squared | 0.714 | 0.522 | 0.619 | 0.510 | 0.669 | 0.601 |

**Figures**

**Figure 1.** *China's Pension System Reforms*

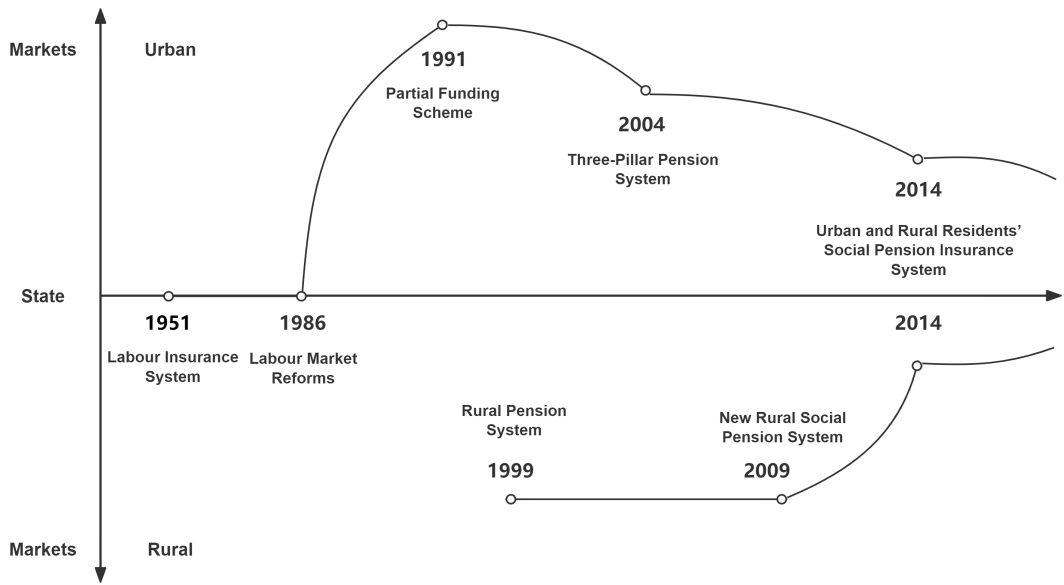

**Figure 2.** *Rural MP Incidence across  k  cutoff*

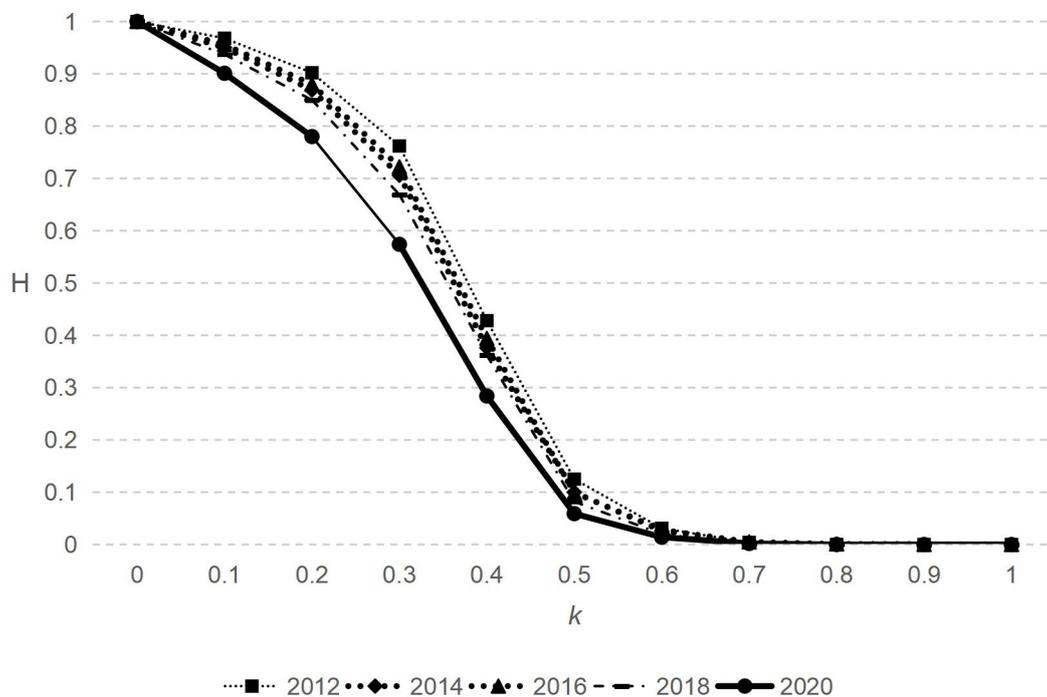

**Figure 3**. *Urban MP Incidence across  k   cutoff*

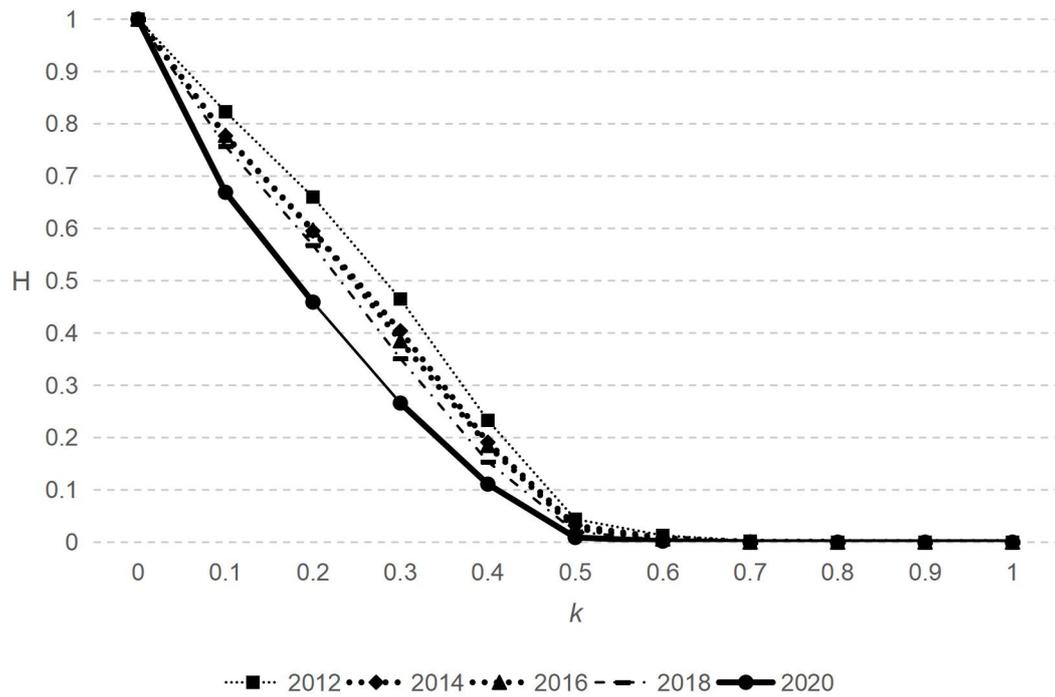

**Figure 4**. *Pension System within Sustainable Livelihoods Framework*

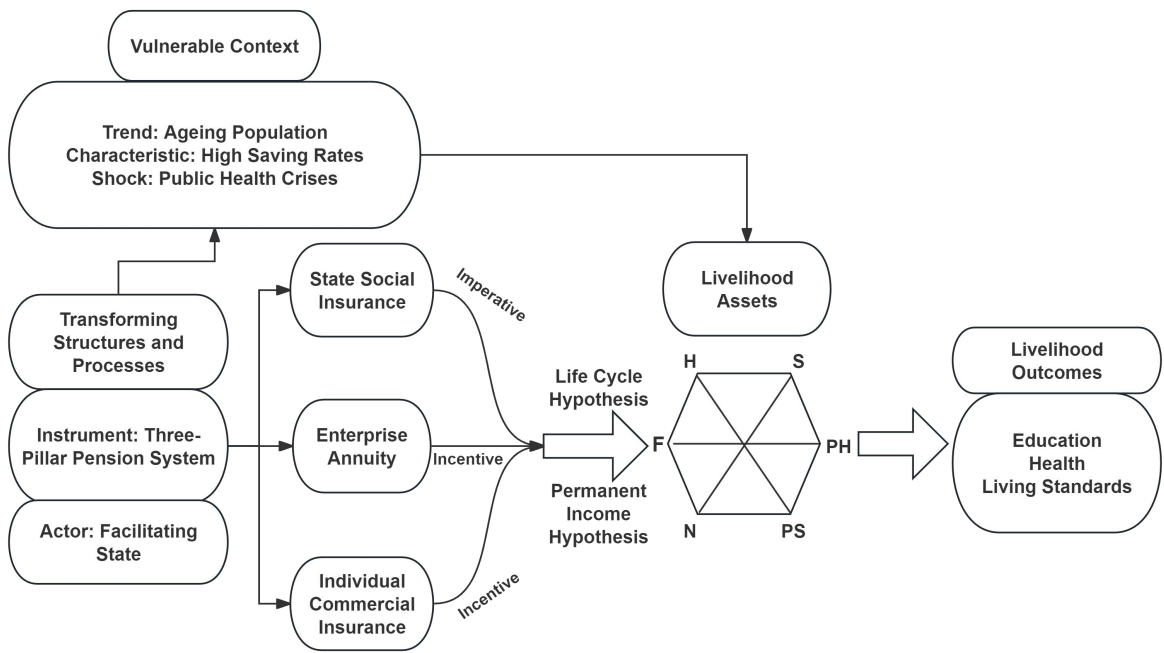

**Appendices**

**Appendant Notes.** *MP Dimensions and Indicators with Income Involved*

Poverty in China was often measured by individual income in research. However, real-world poverty, conceptually multidimensional, entails a nuanced understanding of disparity and deprivation concerning complex dimensions around the household level. MP entails specific dimensions of human development in impoverished settings (UNDP, 2010), corresponding to the human basic needs from a capability approach. Sen (1999) argues that the deprivation and disparity in these needs and capabilities cannot be measured merely by income. Fortunately, Alkire and Foster (2011) shift the measurement. The Alkire-Foster approach selects the dimensions of health, education, and living standards, which is pervasive in MP studies on anti-poverty policy evaluation (Alkire & Santos, 2014), spotlighting diverse manifestations of household poverty. Compared to unidimensional income-focused poverty, the Alkire-Foster approach springs research on poverty conditions of both urban and rural households (Alkire & Fang, 2019).

First, the values for each dimension are established. $x_{ijq}$ represents the value of indicator $q$ for the dimension $j$ of household $i$, $q = 1, 2, ..., s$. $w_{jq}$ is the weight of $q$ within $j$. Thus, the value for the $j$ of $i$ is $y_{ij} = \sum_{q=1}^{s} x_{ijq} w_{jq}$. $M^{n,d}$ is the household dimension matrix, where the element $y \in M^{n,d}$ means the value for the $d\text{-}th$ dimension of the $n\text{-}th$ household across $n$ households on $d$ dimensions. Each $y_{ij}$ in $y$ signifies the value of the $d\text{-}th$ dimension for the $n\text{-}th$ household; a row vector $y_i = (y_{i1}, y_{i2}, ..., y_{id})$ includes all the dimension values for $n\text{-}th$. Similarly, a column vector $y_j = (y_{1j}, y_{2j}, ..., y_{nj})$ implies the value distribution for $d\text{-}th$.

The second is MP identification. Here $z = (z_1, z_2, ..., z_d)$ is the deprivation cutoff matrix, where $z_j (z_j > 0)$ denotes the deprivation cutoff in the $j\text{-}th$ dimension; $g = [g_{ij}]$ is the deprivation matrix, $c = (c_1, c_2, ..., c_d)$ the deprivation count, reflecting the breadth of individual deprivation. The $b$ means the set of deprived dimensions, with $k$ being the MP cutoff. By comparing elements $y_{ij}$ within the household dimension matrix to elements $z_j$ within the deprivation cutoff matrix, the household dimension matrix $M^{n,d}$ is transformed into the deprivation matrix $g_{ij}$:

$$g_{ij} = \begin{cases} 1, & y_{ij} < z_j \\ 0, & y_{ij} \geq z_j \end{cases}$$

By combining the weights of each dimension and the cutoff, it shifts from the unidimensional to our multidimensional measurement:

$$D_i = \sum_{j=1}^{d} g_{ij} w_j$$

$$I(k) = \begin{cases} 1, & D_i \geq k \\ 0, & D_i < k \end{cases}$$

$D_i$ is the MP index for the $i\text{-}th$ household, and $I(k)$ identifies if the household is in the MP status. Additionally, our measurement equally weights each dimension following the UNDP and the OPHI recognised in most MP research. The third step is to sum up the household MP. Alkire and Foster (2011) revise the approach with the MP index $M_0$, the average deprivation share $A$, and the poverty incidence $H$:

$$M_0 = \sum_{i=1}^{n} c_i(b) / nd$$

$$C_j = M_{0j} / M_0 = (q_j \times w_j) / (n \times M_0)$$

$$H = q / n$$

$n$ is the number of households, $q$ signifies the count of the individuals in the MP status when the cutoff of deprivation is set at $k$. The $c_i(b)$ is indicative of the deprivation count, which is determined by comparing the deprived dimension $b$ against the MP cutoff $k$. It is clear that the MP index is determined by the average deprivation share and the poverty incidence. The index $M_0$ is determined by the product of the average deprivation share $A$ and the poverty headcount ratio $H$, such that $M_0 = H \times A$. The fourth is to dissect MP into different dimensions. The expressions of the MP index according to dimensions are as follows:

$$M_{0j} = (q_j \times w_j) / n$$

$$C_j = M_{0j} / M_0 = (q_j \times w_j) / (n \times M_0)$$

In this framework, $M_{0j}$ represents the poverty contribution across dimensions, $q_j$ denotes the poverty headcount ratio for dimension $j$, and $C_j$ signifies the contribution rate of dimension $j$ or region.

**Appendant Table 1.** *MP Dimensions and Indicators with Income Involved*

| Dimension | Indicator | Explanation | Weight |
|---|---|---|---|
| Education | Average Years of Education | Whether the average years of education for members over 16 was less than 9 | 1/8 |
| | School Enrollment | Proportion of children (aged 6-16) who dropped out of school | 1/8 |
| Health | BMI | Whether more than 50% of the adults in the household had a BMI not within the range of 18.5-25 | 1/16 |
| | Hospitalisation | Whether someone in the household was hospitalised in the past year | 1/16 |
| | Self-assessed Health | Whether someone in the household considered themselves to be in very poor health | 1/16 |
| | Health Insurance | Whether someone in the household was not covered by health insurance | 1/16 |
| Living Standards | Cooking Fuel | Whether the household had no access to clean fuel for cooking | 1/20 |
| | Water Access | Whether the household had no access to tap water, bottled water, pure water, or filtered water | 1/20 |
| | Housing Space | Whether the per capita housing area of the household was less than 12 square metres | 1/20 |
| | Electricity Access | Whether the household had no access to electricity | 1/20 |
| | Durable Goods Value | Whether the total value of durable goods in the household was less than 1000 CNY | 1/20 |
| Income | Per Capita Net Income | Whether the per capita net income of the household (in 2011 prices) was less than 2300 CNY | 1/4 |

**Appendant Table 2.** *MP Incidence with Income Involved*

| Heterogeneity | Category | Year | k=0.33 |
|---|---|---|---|
| | | 2012 | 0.302 |
| | | 2014 | 0.259 |
| | H | 2016 | 0.209 |
| | | 2018 | 0.169 |
| | | 2020 | 0.122 |
| | | 2012 | 0.454 |
| | | 2014 | 0.456 |
| All Households | A | 2016 | 0.435 |
| | | 2018 | 0.420 |
| | | 2020 | 0.418 |
| | | 2012 | 0.137 |
| | | 2014 | 0.118 |
| | $M_0$ | 2016 | 0.091 |
| | | 2018 | 0.071 |
| | | 2020 | 0.051 |

| Heterogeneity | Category | Year | k=0.33 | Heterogeneity | Category | Year | k=0.33 |
|---|---|---|---|---|---|---|---|
| | | 2012 | 0.403 | | | 2012 | 0.186 |
| | | 2014 | 0.364 | | | 2014 | 0.152 |
| | H | 2016 | 0.314 | | H | 2016 | 0.11 |
| | | 2018 | 0.265 | | | 2018 | 0.082 |
| | | 2020 | 0.211 | | | 2020 | 0.054 |
| | | 2012 | 0.457 | | | 2012 | 0.446 |
| | | 2014 | 0.459 | | | 2014 | 0.447 |
| Rural Households | A | 2016 | 0.439 | Urban Households | A | 2016 | 0.427 |
| | | 2018 | 0.426 | | | 2018 | 0.415 |
| | | 2020 | 0.422 | | | 2020 | 0.407 |
| | | 2012 | 0.184 | | | 2012 | 0.083 |
| | | 2014 | 0.167 | | | 2014 | 0.068 |
| | $M_0$ | 2016 | 0.138 | | $M_0$ | 2016 | 0.047 |
| | | 2018 | 0.113 | | | 2018 | 0.034 |
| | | 2020 | 0.089 | | | 2020 | 0.022 |

**Appendant Table 3.** *Winsorisation Checks*

| VARIABLES | (1) TPPS | (2) IFMP | (3) TPPS_1 | (4) TPPS_2 | (5) TPPS_3 | (6) IFMP |
|---|---|---|---|---|---|---|
| TPPS | | -0.611*** | | | | |
| | | (0.063) | | | | |
| BartikTPPS | 0.784*** | | | | | |
| | (0.018) | | | | | |
| TPPS_1 | | | | | | -0.255*** |
| | | | | | | (0.067) |
| TPPS_2 | | | | | | -3.312*** |
| | | | | | | (0.227) |
| TPPS_3 | | | | | | -0.463* |
| | | | | | | (0.275) |
| BartikTPPS_1 | | | 0.780*** | | | |
| | | | (0.016) | | | |
| BartikTPPS_2 | | | | 0.531*** | | |
| | | | | (0.033) | | |
| BartikTPPS_3 | | | | | 1.396*** | |
| | | | | | (0.085) | |
| Controls | Y | Y | Y | Y | Y | Y |
| TWFE | Y | Y | Y | Y | Y | Y |
| Observations | 39,020 | 39,020 | 39,020 | 39,020 | 39,020 | 39,020 |
| atanhrho | 0.229*** | - | 0.083*** | 0.668*** | 0.091* | - |

Notes: The survey data might encounter extreme value distributions. Since the impact of extreme values on econometric results, our tests apply the winsorisation to controls and mechanisms. Values above the 99% percentile are limited with the 99% percentile value, and those less than the first percentile are set as the first percentile one. The results basically remain consistent.

**Appendant Table 4.** *Timeframe Checks*

| VARIABLES | (1) TPPS | (2) IFMP | (3) TPPS_1 | (4) TPPS_2 | (5) TPPS_3 | (6) IFMP |
|---|---|---|---|---|---|---|
| TPPS | | -0.586*** | | | | |
| | | (0.067) | | | | |
| BartikTPPS | 0.804*** | | | | | |
| | (0.019) | | | | | |
| TPPS_1 | | | | | | -0.244*** |
| | | | | | | (0.073) |
| TPPS_2 | | | | | | -3.065*** |
| | | | | | | (0.254) |
| TPPS_3 | | | | | | -0.984*** |
| | | | | | | (0.284) |
| BartikTPPS_1 | | | 0.776*** | | | |
| | | | (0.017) | | | |
| BartikTPPS_2 | | | | 0.584*** | | |
| | | | | (0.039) | | |
| BartikTPPS_3 | | | | | 1.343*** | |
| | | | | | (0.096) | |
| Controls | Y | Y | Y | Y | Y | Y |
| TWFE | Y | Y | Y | Y | Y | Y |
| Observations | 31,602 | 31,602 | 31,602 | 31,602 | 31,602 | 31,602 |
| atanhrho | 0.214*** | - | 0.079*** | 0.588*** | 0.177*** | - |

Notes: The year 2020 is a rather unique year. Impacted by the global health crisis, the sample size of CFPS significantly lessens, and the household MP vulnerability was influenced by shocks. Here the data from 2020 are excluded. The results basically remain consistent.

**Appendant Table 5.** *Cluster Robust Standard Error Checks*

| VARIABLES | (1) TPPS | (2) IFMP | (3) TPPS_1 | (4) TPPS_2 | (5) TPPS_3 | (6) IFMP |
|---|---|---|---|---|---|---|
| TPPS | | -0.598*** | | | | |
| | | (0.096) | | | | |
| BartikTPPS | 0.783*** | | | | | |
| | (0.025) | | | | | |
| TPPS_1 | | | | | | -0.206** |
| | | | | | | (0.100) |
| TPPS_2 | | | | | | -3.261*** |
| | | | | | | (0.287) |
| TPPS_3 | | | | | | -0.952*** |
| | | | | | | (0.306) |
| BartikTPPS_1 | | | 0.760*** | | | |
| | | | (0.023) | | | |
| BartikTPPS_2 | | | | 0.559*** | | |
| | | | | (0.043) | | |
| BartikTPPS_3 | | | | | 1.421*** | |
| | | | | | (0.094) | |
| Controls | Y | Y | Y | Y | Y | Y |
| TWFE | Y | Y | Y | Y | Y | Y |
| Observations | 36,923 | 36,923 | 36,923 | 36,923 | 36,923 | 36,923 |
| atanhrho | 0.220*** | - | 0.0659 | 0.644*** | 0.173*** | - |

Notes: Since the potential intracluster correlation of household data based on position of residence, clustering at the level of community offers a more accurate reflection. Here robust standard errors are clustered at the community level. The results basically remain consistent.

**Appendant Table 6.** *Response Variable Checks*

**Section A.** *MPI with Income Involved*

| VARIABLES | (1) TPPS | (2) IFMP | (3) TPPS_1 | (4) TPPS_2 | (5) TPPS_3 | (6) IFMP |
|---|---|---|---|---|---|---|
| TPPS | | -0.674*** | | | | |
| | | (0.071) | | | | |
| BartikTPPS | 0.787*** | | | | | |
| | (0.017) | | | | | |
| TPPS_1 | | | | | | -0.169** |
| | | | | | | (0.068) |
| TPPS_2 | | | | | | -4.089*** |
| | | | | | | (0.211) |
| TPPS_3 | | | | | | -1.059*** |
| | | | | | | (0.369) |
| BartikTPPS_1 | | | 0.783*** | | | |
| | | | (0.017) | | | |
| BartikTPPS_2 | | | | 0.530*** | | |
| | | | | (0.032) | | |
| BartikTPPS_3 | | | | | 1.390*** | |
| | | | | | (0.082) | |
| Controls | Y | Y | Y | Y | Y | Y |
| TWFE | Y | Y | Y | Y | Y | Y |
| Observations | 39,020 | 39,020 | 39,020 | 39,020 | 39,020 | 39,020 |
| atanhrho | 0.248*** | - | 0.051* | 0.921*** | 0.206*** | - |

Notes: Although the Alkire-Foster approach and the MPI index exclude the income dimension in the multidimensional poverty indices, some arguments highlight the importance of income and incorporate it into the index. Accordingly, here our index includes this dimension to re-calculate the MPI. The results remain consistent.

**Section B.** *MPI with k=0.2*

| VARIABLES | (1) TPPS | (2) IFMP | (3) TPPS_1 | (4) TPPS_2 | (5) TPPS_3 | (6) IFMP |
|---|---|---|---|---|---|---|
| TPPS | | -0.792*** | | | | |
| | | (0.061) | | | | |
| BartikTPPS | 0.787*** | | | | | |
| | (0.017) | | | | | |
| TPPS_1 | | | | | | -0.310*** |
| | | | | | | (0.068) |
| TPPS_2 | | | | | | -3.480*** |
| | | | | | | (0.234) |
| TPPS_3 | | | | | | -0.405* |
| | | | | | | (0.240) |
| BartikTPPS_1 | | | 0.783*** | | | |
| | | | (0.016) | | | |
| BartikTPPS_2 | | | | 0.530*** | | |
| | | | | (0.032) | | |
| BartikTPPS_3 | | | | | 1.390*** | |
| | | | | | (0.082) | |
| Controls | Y | Y | Y | Y | Y | Y |
| TWFE | Y | Y | Y | Y | Y | Y |
| Observations | 39,020 | 39,020 | 39,020 | 39,020 | 39,020 | 39,020 |
| atanhrho | 0.301*** | - | 0.092*** | 0.700*** | 0.088** | - |

Notes: The selection of the cutoff value k might cause variations in results. Although k=0.33 in the main text is widely accepted, here k is adjusted to 0.2 to re-calculate the MPI. The results remain consistent.

**Section C.** *MPI with k=0.4*

| VARIABLES | (1)<br>TPPS | (2)<br>IFMP | (3)<br>TPPS_1 | (4)<br>TPPS_2 | (5)<br>TPPS_3 | (6)<br>IFMP |
|---|---|---|---|---|---|---|
| TPPS | | -0.525*** | | | | |
| | | (0.069) | | | | |
| BartikTPPS | 0.787*** | | | | | |
| | (0.017) | | | | | |
| TPPS_1 | | | | | | -0.197*** |
| | | | | | | (0.071) |
| TPPS_2 | | | | | | -3.188*** |
| | | | | | | (0.265) |
| TPPS_3 | | | | | | -0.627* |
| | | | | | | (0.364) |
| BartikTPPS_1 | | | 0.783*** | | | |
| | | | (0.016) | | | |
| BartikTPPS_2 | | | | 0.530*** | | |
| | | | | (0.032) | | |
| BartikTPPS_3 | | | | | 1.390*** | |
| | | | | | (0.082) | |
| Controls | Y | Y | Y | Y | Y | Y |
| TWFE | Y | Y | Y | Y | Y | Y |
| Observations | 39,020 | 39,020 | 39,020 | 39,020 | 39,020 | 39,020 |
| atanhrho | 0.199*** | - | 0.067** | 0.636*** | 0.117* | - |

Notes: Here k is adjusted to 0.4 to re-calculate the MPI. The results remain consistent.

**Appendant Table 7.** *Explanatory Variable Checks*

| VARIABLES | (1) TPPS | (2) IFMP | (3) TPPS_1 | (4) TPPS_2 | (5) TPPS_3 | (6) IFMP |
|---|---|---|---|---|---|---|
| IntensityTPPS | | -0.610*** | | | | |
| | | (0.062) | | | | |
| BartikTPPS | 0.787*** | | | | | |
| | (0.017) | | | | | |
| IntensityTPPS_1 | | | | | | -0.313*** |
| | | | | | | (0.071) |
| IntensityTPPS_2 | | | | | | -5.843*** |
| | | | | | | (0.458) |
| IntensityTPPS_3 | | | | | | -1.098** |
| | | | | | | (0.497) |
| BartikTPPS_1 | | | 0.754*** | | | |
| | | | (0.012) | | | |
| BartikTPPS_2 | | | | 0.353*** | | |
| | | | | (0.027) | | |
| BartikTPPS_3 | | | | | 0.912*** | |
| | | | | | (0.041) | |
| Controls | Y | Y | Y | Y | Y | Y |
| TWFE | Y | Y | Y | Y | Y | Y |
| Observations | 39,020 | 39,020 | 39,020 | 39,020 | 39,020 | 39,020 |
| atanhrho | 0.229*** | - | 0.025 | 0.532*** | 0.087** | - |

Notes: The main text assigns a household a value of 1 for the number of pillars if at least one member participates in any of them, without considering the intensity of participation in a single pillar. By substituting the explanatory variable with TPPS participation rate, calculated as the number of participants in a single pillar divided by the total number of household members, the intensity of household participation in each pillar is better characterised. The results remain consistent after this substitution of the explanatory variable.

**Appendant Table 8.** *Control Variable Checks*

| VARIABLES | (1) TPPS | (2) IFMP | (3) TPPS_1 | (4) TPPS_2 | (5) TPPS_3 | (6) IFMP |
|---|---|---|---|---|---|---|
| TPPS | | -0.607*** | | | | |
| | | (0.062) | | | | |
| BartikTPPS | 0.786*** | | | | | |
| | (0.017) | | | | | |
| TPPS_1 | | | | | | -0.245*** |
| | | | | | | (0.065) |
| TPPS_2 | | | | | | -3.352*** |
| | | | | | | (0.226) |
| TPPS_3 | | | | | | -0.509* |
| | | | | | | (0.273) |
| BartikTPPS_1 | | | 0.782*** | | | |
| | | | (0.016) | | | |
| BartikTPPS_2 | | | | 0.529*** | | |
| | | | | (0.032) | | |
| BartikTPPS_3 | | | | | 1.388*** | |
| | | | | | (0.082) | |
| Controls | Y | Y | Y | Y | Y | Y |
| TWFE | Y | Y | Y | Y | Y | Y |
| Observations | 39,020 | 39,020 | 39,020 | 39,020 | 39,020 | 39,020 |
| atanhrho | 0.228*** | - | 0.080*** | 0.675*** | 0.098** | - |

Notes: MP and participation in the pension system are influenced by macroeconomic factors. However, the controls are mainly at the household and individual levels, without accounting for the macroeconomic factors, which might cause biased estimation results. Building upon the baseline regressions, the control variables at the regional level are introduced, including regional GDP, the value-added index of the tertiary sector, the number of regular higher education institutions, the volume of library collections, the number of medical and healthcare institutions, the number of beds in healthcare institutions, and the count of enterprises involved in the production and supply of electricity, thermal power, gas, and water. The macro-level variables are expected to influence the MP of households. The results remain consistent.

**Appendant Table 9.** *Propensity Score Matching Checks*

| VARIABLES | (1) TPPS | (2) IFMP | (3) TPPS_1 | (4) TPPS_2 | (5) TPPS_3 | (6) IFMP |
|---|---|---|---|---|---|---|
| TPPS | | -0.559*** | | | | |
| | | (0.067) | | | | |
| BartikTPPS | 0.818*** | | | | | |
| | (0.020) | | | | | |
| TPPS_1 | | | | | | -0.229*** |
| | | | | | | (0.070) |
| TPPS_2 | | | | | | -3.420*** |
| | | | | | | (0.267) |
| TPPS_3 | | | | | | -0.647** |
| | | | | | | (0.297) |
| BartikTPPS_1 | | | 0.819*** | | | |
| | | | (0.018) | | | |
| BartikTPPS_2 | | | | 0.513*** | | |
| | | | | (0.036) | | |
| BartikTPPS_3 | | | | | 1.338*** | |
| | | | | | (0.085) | |
| Controls | Y | Y | Y | Y | Y | Y |
| TWFE | Y | Y | Y | Y | Y | Y |
| Observations | 29,367 | 29,367 | 29,367 | 29,367 | 29,367 | 29,367 |
| atanhrho | 0.214*** | - | 0.083*** | 0.665*** | 0.114** | - |

Notes: Since household characteristics influence MP status, the propensity score matching is employed to minimise selection bias. The one-to-one matching forms pairs of treatment and control groups based on TPPS participation. Regressing matched samples with similar household characteristics augments the robustness.